\documentclass[preprint,12pt,fleqn]{elsarticle}

\usepackage{amssymb}

\usepackage{amsmath,slashed}          
\usepackage{seceqn}                   %%FRK: IMPORTANT add file seceqn.sty
\def\id{\makebox[0.6ex][l]{$1$}{\rm l}}   %%FRK from KlinkhamerSchimmel2002
\newcommand{\overstarn}{\stackrel{\star}{n}} %%FRK in revtex have \overstar

\journal{}

\begin{document}

\begin{frontmatter}

\title{Anomalous Lorentz and CPT violation\\
from a local Chern--Simons-like term\\
in the effective gauge-field action}

\author{K.J.B. Ghosh}
\ead{kumar.ghosh@kit.edu}

\author{F.R. Klinkhamer\corref{cor1}}
\cortext[cor1]{Corresponding author}
\ead{frans.klinkhamer@kit.edu}
\address{Institute for
Theoretical Physics, Karlsruhe Institute of
Technology (KIT),\\ 76128 Karlsruhe, Germany}

\begin{abstract}
We consider four-dimensional chiral gauge theories defined over
a spacetime manifold with topology $\mathbb{R}^3 \times S^1$
and periodic boundary conditions over the compact dimension.
The effective gauge-field action is calculated for Abelian
$U(1)$ gauge fields
$A_{\mu}(x)$ which depend on all four spacetime
coordinates (including the coordinate $x^{4}\in S^1$ of the compact dimension) and have vanishing components $A_{4}(x)$
(implying trivial holonomies in the 4-direction).
Our calculation shows that the effective gauge-field action contains
a local Chern--Simons-like term which violates Lorentz and CPT invariance.
This result is established perturbatively with a generalized
Pauli--Villars regularization and nonperturbatively with a lattice
regularization based on Ginsparg--Wilson fermions.
\end{abstract}


\end{frontmatter}

\section{Introduction}
\label{sec:introduction}

It has been shown~\cite{Klinkhamer1999}
that chiral gauge theories over a
manifold with an appropriate nontrivial topology necessarily
have an anomalous violation of Lorentz and CPT invariance.
Two direct follow-up papers on this CPT anomaly have appeared in
Refs.~\cite{KlinkhamerNishimura2000,KlinkhamerSchimmel2002}
and a review has been presented in Ref.~\cite{Klinkhamer2005}
which also contains a brief discussion of the well-known CPT theorem
and ways how this theorem can be circumvented.

The existence of the CPT anomaly for four-dimensional gauge
chiral theories over the spacetime manifold
$M=\mathbb{R}^3 \times S^1$ was established
in Refs.~\cite{Klinkhamer1999,KlinkhamerSchimmel2002}
for a special class of background gauge fields,
namely gauge-field configurations which are independent of the compact
coordinate $x^4\in S^1$ and have a vanishing component $A_4$.
The question arises how the anomaly manifests itself for more
general gauge-field configurations which have a nontrivial
dependence on the compact $x^4$ coordinate.

It will be shown,
in the present article, that the anomaly manifests itself
by a \emph{local}
Chern--Simons-like term in the effective gauge-field action
and this term is known to violate Lorentz
and CPT invariance~\cite{ChadhaNielsen1982,%
CarrollFieldJackiw1990,ColladayKostelecky1998}.
Our result will be established with two regularization methods,
an extended version of
the generalized Pauli--Villars regularization~\cite{FrolovSlavnov1993}
for a perturbative calculation
and the lattice regularization based on Ginsparg--Wilson
fermions~\cite{GinspargWilson1982,Neuberger1998a,Neuberger1998b,%
Luscher1998a,Luscher1998b} for a nonperturbative calculation.

The outline of this article is as follows.
In Sec.~\ref{sec:Setup-problem}, we describe
the theo\-re\-ti\-cal setup of the problem and establish our notation.
As said, the calculation will be done both perturbatively and
nonperturbatively, with appropriate regularization methods.

In Sec.~\ref{sec:Perturbative-approach}, we establish Lorentz and CPT violation with a perturbative approach.
In Sec.~\ref{subsec:Theory-and-regularization}, we start from
the effective gauge-field action for a left-handed chiral fermion.
This effective action is then perturbatively expanded
and rendered finite with an extended version of
the generalized Pauli--Villars regularization.
In Sec.~\ref{subsec:Calculation-pert}, we perform, for an Abelian
$U(1)$ gauge group, the one-loop calculation of the
effective gauge-field action to quadratic order and
obtain a local Chern--Simons-like term.
In Sec.~\ref{subsec:Lorentz and CPT violation}, we show explicitly
that the calculated Chern--Simons-like term violates Lorentz
and CPT invariance in four spacetime dimensions.

In Sec.~\ref{sec:Nonperturbative-approach}, we establish
the existence of Lorentz and CPT violation with a nonperturbative approach.
In Sec.~\ref{subsec:Lattice-setup}, we recall the lattice setup and introduce some further notation.
In Sec.~\ref{subsec:Chiral fermions on the lattice}, we review
chiral $U(1)$ gauge theory on the lattice.
The fermion action on a regular hypercubic lattice is written down and the integration measure is defined. The action of the discrete transformations on the link variable is also given.
In Sec.~\ref{subsec:Effective action and CPT transformation}, we discuss
the effective gauge-field action on the lattice and its behavior under a CPT transformation.
In Sec.~\ref{subsec:CPT anomaly}, we show that the effective
action is not invariant under CPT transformation,
considering both relevant cases (an odd or even integer $N\equiv L/a$,
with $L$ the length of the $x^4$ circle and $a$ the lattice spacing).
In Sec.~\ref{subsec:Continuum-limit}, we calculate the expression for the CPT-anomaly in the continuum limit ($a\to 0$).

In Sec.~\ref{sec:Discussion}, we highlight some important
points of our calculations.
In Sec.~\ref{sec:Conclusion}, finally, we offer some concluding remarks.

The present article is, by necessity, rather technical. A first impression
can be obtained from Secs.~\ref{sec:Setup-problem},
\ref{subsec:Lorentz and CPT violation}, and \ref{sec:Conclusion}.

%%\newpage%%tmp
\section{Setup of the problem}
\label{sec:Setup-problem}

The chiral gauge theory to be considered is defined over the
 following four-dimensional spacetime manifold:
\begin{subequations}\label{eq:M-x12range-x4range}
\begin{eqnarray}
\label{eq:M}
M = \mathbb{R}^3 \times S^1   \,,
\end{eqnarray}
with noncompact coordinates
\begin{eqnarray}
x^1 , x^2, x^3\in \mathbb{R}  \,,
\end{eqnarray}
and compact coordinate
\begin{eqnarray}
x^4 \in [0 , L]  \,.
\end{eqnarray}
\end{subequations}
Initially, the spacetime metric is taken to be the Euclidean flat metric,
\begin{equation}\label{eq:Euclidean-flat-metric}
g_{\mu\nu}(x)=[\text{diag}(1,\, 1,\, 1,\, 1)]_{\mu\nu}\,.
\end{equation}
At the end of the calculation, we shall make the Wick rotation
from Euclidean metric signature to Lorentzian metric signature,
with $x^{1}$ or $x^{2}$ or $x^{3}$ (but not $x^4$)
taken to correspond to the time coordinate $t$.

We are considering chiral gauge theories that are free
of gauge anomalies. Specifically, we take the
chiral gauge theory with the following non-Abelian gauge group
and representation of left-handed fermions:
\begin{subequations}
\label{eq:SO10theory-G-RL}
\begin{eqnarray}
\label{eq:SO10theory-G}
\hspace*{-5mm}
G &=& SO(10)\,,
\\[2mm]
\label{eq:SO10theory-RL}
\hspace*{-5mm}
R_L &=& 3 \times [\mathbf{16}]\,,
\end{eqnarray}
\end{subequations}
which contains the $SU(3) \times SU(2) \times U(1)$
Standard Model with 3 families of fermions (and three
singlet left-handed antineutrinos).

Most of our calculations are, however, performed for
a chiral $U(1)$ gauge theory
consisting of a single gauge boson $A$ and 48 left-handed
fermions with $U(1)$ charges $q_f$, for $f=1, \ldots , 48$.
Specifically,  the  Abelian gauge group and the left-handed fermion representation (i.e., the set of left-handed charges $q_f$
in units of $e$, the absolute value of the
electron charge) are given by:
\begin{subequations}\label{eq:U1theory-G-RL}
\begin{eqnarray}
\label{eq:U1theory-G}
\hspace*{-5mm}
G  &=& U(1) \,,
\\[2mm]
\label{eq:U1theory-RL}
\hspace*{-5mm}
R_L&=& 3\times \Bigg[
  6\times \left( \frac{1}{3}\right)
+ 3\times \left(-\frac{4}{3}\right)
+ 3\times \left( \frac{2}{3}\right)
\nonumber\\[1mm] &&
+ 2\times \left(-1\right)
+ 1\times \left(2\right)
+ 1\times \left(0\right) \Bigg]\,.
\end{eqnarray}
\end{subequations}
This particular chiral $U(1)$ gauge theory can be embedded in
the $SU(2)\times U(1)$ electroweak theory of the Standard Model
with $U(1)$ hypercharge $Y\equiv 2\,Q-2\,T_3$ (the electron has
charge $Q=-e$ and the positron has $Q=+e$.)
The further embedding in the ``safe'' $SO(10)$ group
with left-handed representation \eqref{eq:SO10theory-RL}
explains why the perturbative  gauge anomalies cancel out in the chiral
$U(1)$ gauge theory considered,
\begin{equation}\label{eq:sum-charges-cube-Sum-F}
\sum_{f=1}^{48}\; (q_f)^3 =0\,,
\end{equation}
for the charges $q_f$ as given by \eqref{eq:U1theory-RL}.
For later use, we also give another sum:
\begin{subequations}\label{eq:sum-charges-square-Sum-F}
\begin{eqnarray}
\label{eq:sum-charges-square-Sum}
\sum_{f=1}^{48}\; (q_f)^2 &=& F\,e^2\,,
\\[2mm]
\label{eq:sum-charges-square-F}
F &=& 3 \times \left[\frac{40}{3}\right] = 40\,.
\end{eqnarray}
\end{subequations}
Other chiral $U(1)$ gauge theories give, in general,
a different value for the numerical factor $F$.

The gauge and fermion fields
are assumed to be periodic in the $x^4$ coordinate,
\begin{subequations}\label{eq:periodic-bcs}
\begin{eqnarray}
A_\mu(\vec{x},\, x^4+L)
&=&
A_\mu(\vec{x},\, x^4)   \,,
\\[2mm]
\psi(\vec{x},\, x^4+L)
&=&
\psi(\vec{x},\, x^4)  \,,
\\[2mm]
\overline{\psi}(\vec{x},\, x^4+L)
&=&
\overline{\psi}(\vec{x},\, x^4)   \,,
\end{eqnarray}
\end{subequations}
with
\begin{equation}\label{eq:def-vec-x}
\vec{x}\equiv (x^1,\,x^2,\, x^3)\,.
\end{equation}
  Another assumption about the gauge fields is as follows:
\begin{subequations}\label{eq:zero-A4-condition}
\begin{eqnarray}
A_i(x) &=& A_i(\vec{x},\, x^4) \,, \;\;
\text{for}\;\; i=1,\,2,\,3 \,,
\\[2mm]
A_4 (x) &=& 0   \,.
\end{eqnarray}
\end{subequations}
Such gauge fields can be obtained by a gauge transformation if
the original gauge fields with $A_4 \ne 0$ have trivial holonomies,
\begin{equation}\label{eq:trivial-holonomy-A}
h_4(\vec{x}) \equiv
\exp \left[\int_{0}^{L}\, d x^4\, A_4(\vec{x},\, x^4) \right]
=1 \,.
\end{equation}
This Abelian holonomy $h_4(\vec{x})$ is a gauge-invariant quantity
(see the last paragraph of Sec.~\ref{subsec:Calculation-pert}).

The background gauge fields $A_i$ are considered
to have local support in $\mathbb{R}^3$.
Specifically, take a ball $B^3 \in \mathbb{R}^3$
with a large fixed radius $R$.
The gauge fields $A_i(x)$, for $i=1,\,2,\,3$,
are assumed to vanish on the boundary of
the ball and outside of it,
\begin{equation}\label{eq:ball-condition-A}
A_i(\vec{x},\, x^4)= 0\,,
\;\;\text{for}\;\;
|\vec{x}|^2 \equiv (x^1)^2+(x^2)^2+(x^3)^2\geq R^2\,.
\end{equation}
In general, Latin spacetime indices $i,j,k, l$, etc. run over the coordinate labels $1, 2, 3$,
and Greek spacetime indices $\mu, \nu,\, \rho$, etc. over
the labels $1,\, 2,\, 3,\, 4$. Repeated coordinate (and internal) indices are summed over. Throughout, natural units are used with $\hbar=c=1$.

The problem, now, is to investigate, for the setup considered,
the invariance of the effective gauge-field action $\Gamma[A]$
under Lorentz and CPT transformations.
In Secs.~\ref{sec:Perturbative-approach} and
\ref{sec:Nonperturbative-approach}, the
effective action $\Gamma[A]$ is calculated
by integrating out the fermions using, respectively,
a perturbative and a nonperturbative method.
The CPT anomaly is then established if we can show
that this effective action changes under a CPT transformation of the
background gauge field, $\Gamma[A^\text{CPT}] \ne \Gamma[A]$.

The actual calculation of Sec.~\ref{sec:Perturbative-approach}
is performed first
for a single left-handed fermion $\psi$ with unit $U(1)$ charge, $q=e$.
Only the final result \eqref{eq:mathcalT-anom} is extended to
all chiral fermions of the theory \eqref{eq:U1theory-G-RL}.
The same procedure is followed in
Sec.~\ref{sec:Nonperturbative-approach}.

%%\newpage%%tmp
\section{Perturbative approach}
\label{sec:Perturbative-approach}

\subsection{Theory and regularization}
\label{subsec:Theory-and-regularization}

Let us start with the action of a left-handed chiral fermion,
\begin{eqnarray}
S \left[\,\overline{\psi} ,\, \psi ,\, A \right] &=& ~
\int_M d^{4} x  ~
\mathcal{L}\left[\,\overline{\psi_L} , \psi_L, A \right] \nonumber \\
&=& \int_M d^{4} x ~i ~ \overline{\psi_L} \, \gamma ^\mu
( {\partial _\mu} + e\,A_\mu )~ \psi_L \,,
\label{eq:LH-fermion-action}
\end{eqnarray}
where $A_\mu$ is the anti-Hermitian $U(1)$ gauge field,
$e$ the dimensionless electric charge of the fermion $\psi$,
and $\psi_L \equiv\frac{1}{2} (1+ \gamma_5)\, \psi$
the left-handed projection of the four-component Dirac spinor $\psi$.
The $\gamma^\mu$ are the $4\times 4$ Dirac matrices
and $\overline{\psi}\equiv \psi^\dagger\, \gamma ^4$.
The Hermitian chirality matrix $\gamma_5$ has
$\{\gamma_5,\, \gamma^\mu\}=0$ and $(\gamma_5)^2= \id_{\,4}$.

In this article, we set out to calculate the effective action of the gauge
fields for the setup as described in Sec.~\ref{sec:Setup-problem}.
In the vacuum, there are virtual fermion-antifermion pairs which interact with the classical background gauge field. The effective action $\Gamma[A]$ is a functional which takes these interactions into account.
Incidentally, the functional $\Gamma[ A ]$ considered here
is not the complete effective action as there are also
contributions from the photonic sector such as the classical
Maxwell term, but our focus is solely on the contributions
of the virtual fermions.

In Feynman's Euclidean path integral formalism, the functional $\Gamma[A]$ is obtained by integrating out the fermionic degrees of freedom,
\begin{equation}
\label{eq:exp-Gamma-A}
\exp(-\Gamma[ A ])
= \int \mathcal{D} \overline{\psi_L}(x) \mathcal{D} \psi_L(x) ~
\exp\left(-\int_M d^{4}x ~
\mathcal{L}\left[\,\overline{\psi_L} , \psi_L, A\right] \right),
\end{equation}
which, loosely speaking, equals the root of the determinant
of the operator $\gamma ^\mu ( {\partial _\mu} +e\, A_\mu )$.
This operator has, however, an unbounded spectrum,
so that the determinant is infinite .
The expression \eqref{eq:exp-Gamma-A} thus needs to be regularized.

Finding a manifestly gauge-invariant regularization is not straightforward. One possibility
is given by the generalized Pauli--Villars regularization
as discussed by Frolov and Slavnov \cite{FrolovSlavnov1993}, which
involves an infinite set of bosonic
and fermionic Pauli--Villars-type fields $\Psi_{s}$,
for $s \in \mathbb{Z}/\{0\}$, with standard (Lorentz-invariant)
Dirac-type mass terms $m_{s}\,\overline{\Psi}_{s}\,\Psi_{s}$.
We will, however, extend this regularization,
in order to be sensitive to anomalous Lorentz violation.
In fact, we will introduce  another infinite set of bosonic
and fermionic Pauli--Villars-type fields $\psi_r$,
for $r \in \mathbb{Z}/\{0\}$,
with Lorentz-violating mass terms $M_r\,\psi_r^\dagger\, \psi_r$.

Specifically, the regularized Lagrange density for
the chiral $U(1)$ gauge theory
including both infinite sets of Pauli--Villars-type fields
reads as follows:
\begin{eqnarray}
\label{eq:full-PV-type-lagrangian}
\hspace*{-10mm}
\mathcal{L}_{\text{full\;reg.\;th.}} &=&
\mathcal{L}_{\text{chiral}}+ \mathcal{L}_\text{LI-gen-PV}
+ \mathcal{L}_\text{LV-gen-PV}
\nonumber \\[1mm]
\hspace*{-10mm}
 &=&
i~ \overline{\psi_{0}}(x) ~  \gamma^\mu \left(\partial_\mu
+ e\,A_\mu \right) \psi_{0}(x)
\nonumber \\[1mm]
\hspace*{-10mm}
&&
+
\sum_{s\ne 0}
\left[
i~ \overline{\Psi}_{s}(x) ~  \gamma^\mu \left(\partial_\mu
+ e\,A_\mu \right) \Psi_{s}(x)
-m_{s}\,\overline{\Psi}_{s}(x)\,\Psi_{s}(x)
\right]
\nonumber \\[1mm]
\hspace*{-10mm}
&&
+
\sum_{r\ne 0}
\left[
i~ \overline{\psi_{r}}(x) ~  \gamma^\mu \left(\partial_\mu
+ e\,A_\mu \right) \psi_{r}(x)
-M_r\, \psi_{r}^\dagger (x)\, \psi_{r}(x)
\right]\,,
\end{eqnarray}
with regulator masses,
\begin{subequations}\label{eq:regulator-masses-Ms-Mr-inequality}
\begin{eqnarray}
\label{eq:regulator-masses-Ms}
m_{s} &=& m\,|s|   \,,
\\[2mm]
\label{eq:regulator-masses-Mr}
M_r &=& M\,r^2   \,,
\\[2mm]
\label{eq:regulator-masses-inequality}
M &\gg& m  \,.
\end{eqnarray}
\end{subequations}
The ultraheavy regulator masses $M_r$ violate Lorentz invariance,
but can have effects on the low-energy physics in the case of an
anomaly. The reason for demanding a quadratic $r$-dependence
in  \eqref{eq:regulator-masses-Mr},
compared to the linear $s$-dependence in \eqref{eq:regulator-masses-Ms},
will be explained in Sec.~\ref{subsec:Calculation-pert}.
Strictly speaking, we do not need the inequality
\eqref{eq:regulator-masses-inequality}
for the present calculation, but it has been included,
in order to make sure that possible Lorentz-violating quantum effects
are subdominant compared to Lorentz-invariant quantum effects.

The regulator fields $\Psi_{s}$ in \eqref{eq:full-PV-type-lagrangian}
are unrestricted four-component Dirac fields,
whereas the regulator fields $\psi_r$, including the original
massless field $\psi_0 \equiv \psi_L$, are chiral four-component Dirac fields, obeying the condition
\begin{equation}
\psi_r \equiv  \frac{1}{2}\,(1+ \gamma_5)\, \psi_r
\,,\;\;\text{for}\;\;r \in \mathbb{Z}\,.
\end{equation}
The fields have, moreover, the following Grassmann parities:
\begin{subequations}\label{eq:Grassmann-parity}
\begin{eqnarray}
\label{eq:Grassmann-parity-psi-s}
\varepsilon (\Psi_{s}) &=& (-1)^{s+1}
\,,\;\;\text{for}\;\;s \in \mathbb{Z}/\{0\}\,,
\\[2mm]
\label{eq:Grassmann-parity-psi-r}
\varepsilon (\psi_r) &=& (-1)^{r+1}\,,
\;\;\text{for}\;\;r \in \mathbb{Z}\,.
\end{eqnarray}
\end{subequations}
For the purpose of searching for anomalous Lorentz violation,
we only need to consider the chiral fields $\psi_r$, as will be explained
in Sec.~\ref{subsec:Calculation-pert}.

We now take the Weyl representation of the $4\times 4$
Dirac gamma matrices,
\begin{equation}
\label{eq:Weyl-representation}
\gamma^\mu = \left(\begin{array}{cc}
 0   &  \widetilde{\sigma}^\mu  \\
 \widetilde{\sigma}^{\mu}{}^\dagger   &  0  \\
                              \end{array}\right)\,,
\quad
\gamma_{5} \equiv \gamma^1 \gamma^2 \gamma^3 \gamma^4
=
\left(\begin{array}{cc}
 \id_{\,2}   &  0  \\
 0   &  -\id_{\,2}  \\
                              \end{array}\right)\,,
\end{equation}
with $\widetilde{\sigma}^\mu \equiv (\sigma^m ,\, i\,\id_{\,2})$
in terms of the
$2\times 2$ Pauli spin matrices $\sigma^m$
and the $2\times 2$ identity matrix $\id_{\,2}$.
As said before, $\psi_0$ with $M_0=0$ in (\ref{eq:full-PV-type-lagrangian})
corresponds to the original four-component chiral field
$\psi_L$ and,
for the Weyl representation \eqref{eq:Weyl-representation}
with diagonal $\gamma_{5}$,  can be written as
\begin{equation}
\psi_0 = \left(\begin{array}{c}
                                \xi_0    \\
                                 0  \\
                              \end{array}\right)\,,
\end{equation}
where $\xi_0$ is an anticommuting two-component spinor field.
The $r\neq0$ fields $\psi_r$  in (\ref{eq:full-PV-type-lagrangian})
constitute an infinite set of Pauli--Villars
fields with Grassmann parities \eqref{eq:Grassmann-parity-psi-r}
and regulator masses \eqref{eq:regulator-masses-Mr}.
Each chiral regulator field $\psi_r$ $(r\neq0)$  can also be written as
\begin{equation}
\psi_r = \left(\begin{array}{c}
                                \xi_r    \\
                                 0  \\
                              \end{array}\right)\,,
\end{equation}
with a two-component field $\xi_r$  having the Grassmann parity
(i.e., loop-factor in Feynman diagrams)
\begin{equation}
\varepsilon (\xi_r) = (-1)^{r+1}\,,
\;\;\text{for}\;\;r \in \mathbb{Z}\,.
\label{Grassmann_Parity}
\end{equation}
With the above definitions, the truncated regularized theory
is given by
\begin{eqnarray}
\hspace*{-10mm}
\mathcal{L}_{\text{trunc.\;reg.\;th.}}
&=&
\mathcal{L}_{\text{chiral}}+ \mathcal{L}_\text{LV-gen-PV} \nonumber \\
\hspace*{-10mm}
 &=& \sum_{r=-\infty} ^{\infty}
\Big[
i~ \xi_r^\dagger (x)~\sigma^\mu
\left(\partial_\mu + e\,A_\mu \right) \xi_r(x)
-M_r\, \xi_r^\dagger (x)\, \xi_r(x)
\Big],
\label{eq:trancated-PV-type-lagrangian}
\end{eqnarray}
with $\sigma ^\mu \equiv ( i \sigma^m ,\, \id_{\,2})$
and $M_r$ from \eqref{eq:regulator-masses-Mr}.

In order to prepare for the calculation of the next
subsection, we define
\begin{subequations}\label{eq:3D-gamma-matrices}
\begin{eqnarray}
\widetilde{\gamma}^1 &\equiv& i\,   \sigma ^1=\left(\begin{array}{cc}
                                0   &  i  \\
                                i   &  0  \\
                              \end{array}\right)\,,\quad
\widetilde{\gamma}^2 \equiv i \,  \sigma ^2=\left(\begin{array}{cc}
                                0   &   1 \\
                                -1   &   0 \\
                              \end{array}\right)\,,
\\[2mm]
\widetilde{\gamma}^3 &\equiv& i \, \sigma ^3=\left(\begin{array}{cc}
                                i       &   0 \\
                                0       &   -i  \\
                              \end{array}\right)\,,\quad
\widetilde{\gamma}^4 \equiv   \id_{\,2} =\left(\begin{array}{cc}
                                1   &  0  \\
                                0   &  1  \\
                              \end{array}\right)  \,,
\end{eqnarray}
\end{subequations}
and rewrite the standard Weyl action from
\eqref{eq:trancated-PV-type-lagrangian} as
\begin{equation}\label{eq:I0-3D-gamma-matrices}
\mathcal{I}_0 = \int d^{4}x ~ \mathcal{L}_{\text{chiral}}
= \int d^{4}x~i~
\xi_0^\dagger (x) ~ \widetilde{\gamma}^\mu(\partial_\mu + e\,A_\mu) ~ \xi_0\,,
\end{equation}
where $\xi_0$ is the two-component spinor field.
A similar action holds for the chiral regulator fields
$\xi_r$ $(r\neq0)$,
\begin{eqnarray}
\label{eq:Ireg-3D-gamma-matrices}
\mathcal{I}_\text{reg}
&=&
\int d^{4}x ~ \mathcal{L}_\text{LV-gen-PV}
\nonumber\\[1mm]
&=&
\int d^{4}x~\sum_{r\neq0} ~
\Big[i~\xi_r^\dagger (x) ~ \widetilde{\gamma} ^\mu(\partial_\mu + e\,A_\mu) ~\xi_r - M_r\, \xi_r^\dagger\, \xi_r \Big]\,.
\end{eqnarray}

The $2\times 2$ matrices $\widetilde{\gamma}^\mu$
in \eqref{eq:I0-3D-gamma-matrices} and \eqref{eq:Ireg-3D-gamma-matrices}
obey the following relation:
\begin{equation}\label{eq:3D-gamma-matrices-relation}
\widetilde{\gamma}^i \,\widetilde{\gamma}^j
= \tilde{g}^{ij}\,\id
-  \epsilon^{ijk} \, \widetilde{\gamma}_k\,,
\end{equation}
with the three-dimensional Euclidean flat metric
$\tilde{g}^{ij}=[\text{diag}(-1,-1,-1)]^{ij}$
and  the totally antisymmetric
Levi-Civita symbol $\epsilon^{ijk}$, normalized by $\epsilon^{123}=1$.
From \eqref{eq:3D-gamma-matrices-relation},
we have that the anti-commutator of the $\widetilde{\gamma}^i$
matrices has precisely the same structure as the one
of Dirac matrices in $\mathbb{R}^3$, namely,
$\{ \widetilde{\gamma}^i,\,\widetilde{\gamma}^j \}
= 2\, \tilde{g}^{ij}\,\id$.  This is, in fact, the
reason for using these matrices $\widetilde{\gamma}^\mu$,
as will become clear in Sec.~\ref{subsec:Calculation-pert}.
Note, however, that the matrices $\widetilde{\gamma}^\mu$ do not satisfy
the properties of Dirac gamma matrices in four-dimensional
spacetime,
because $\widetilde{\gamma}^4$ does not anti-commute
with the other $\widetilde{\gamma}^i$ matrices.
In our calculations, we shall  only use
relation (\ref{eq:3D-gamma-matrices-relation}).

For standard Minkowski spacetime without compactification of the $x^4$ coordinate, we expand the gauge field $A_\mu$ in Fourier modes
as follows:
\begin{equation}
A_\mu (x) = \int \frac{d^{4}p}{(2\pi)^4} ~ e^{i p\cdot x}~ {A_\mu} (p),
\end{equation}
and write down the vacuum-polarization kernel
\begin{equation}
\pi^{ij} (p) = \int \frac{d^{4} k}{(2\pi)^4}
\,\text{tr}\, \Big[
\widetilde{\gamma}^i\, S(k)\,\widetilde{\gamma}^j \, S(k+p)\Big]\,.
\end{equation}
In our case, where the $x^4$ coordinate is compactified, we make the following replacements:
\begin{subequations}\label{eq:replacement-d4x-d4p}
\begin{equation}\label{eq:replacement-d4x}
\int d^{4} x \rightarrow \int_0^L d x^4 \int_{\mathbb{R}^3} d^{3} x
\end{equation}
and
\begin{equation}
\int \frac{d^{4}p}{(2\pi)^4} \rightarrow \frac{1}{L} \sum_{n={-\infty}}^{\infty}  \int \frac{d^{3}p}{(2\pi)^3}.
\end{equation}
\end{subequations}
The Fourier expansion of the gauge field $A_\mu$ is now given by
\begin{equation}
A_\mu (x) = \frac{1}{L} \sum_{n={-\infty}}^{\infty} \int \frac{d^{3}p}{(2\pi)^3} ~ e^{2\pi i n x^4/L}~ e^{i \vec{p}\cdot \vec{x}}~ {A_\mu} (p_n),
\end{equation}
with the following definitions:
\begin{subequations}\label{eq:defs-pn-pn2-rhon}
\begin{eqnarray}
p_n  &\equiv&(\vec{p},\, \rho_n )  \,,
\\[2mm]
\rho_n  &\equiv& 2\pi n/L   \,,
\\[2mm]
p_n^2 &\equiv& |\vec{p}|^2+(\rho_n)^2  \,.
\end{eqnarray}
\end{subequations}

%%\newpage%%tmp
\subsection{Calculation}
\label{subsec:Calculation-pert}

The expression for the perturbatively-expanded
effective gauge-field action in three spacetime dimensions
with one compactified coordinate
has been given in Ref.~\cite{AitchisonFoscoZuk1993};
see, in particular,  Eqs.~(22)--(26) of that article.
For the action \eqref{eq:I0-3D-gamma-matrices}
with the replacement \eqref{eq:replacement-d4x},
we have four spacetime dimensions with one compactified coordinate.
Adopting a similar procedure as the one of
Ref.~\cite{AitchisonFoscoZuk1993}, we
write down the physically relevant factor in the perturbatively-expanded effective gauge-field action,
\begin{equation}
\hspace*{-0mm}
\Gamma[A] =  -i\,\frac{e^2}{2}
\frac{1}{L} \sum_{n=-\infty}^\infty \int \frac{d^{3} p}{(2\pi)^3}\,
A _{i} (-p_n)\, \pi^{ij} (p_n)\, A _{j} (p_n)  + O(e^3),
\label{factor1}
\end{equation}
with the unregularized vacuum-polarization kernel
\begin{equation}
\pi^{ij} (p_n)\,\Big|^\text{(unreg.)}
= \frac{1}{L} \sum_{m=-\infty}^\infty \int \frac{d^{3} k}{(2\pi)^3}\, \text{tr}\,
\Big[\widetilde{\gamma}^i\, S(k_m)\,\widetilde{\gamma}^j\, S(k_m+p_n)\Big]. \label{eq:pimunu}
\end{equation}
The propagator $S(k_m)$ is defined as:
\begin{equation}
\label{eq:propagator-S-of-km}
S(k_m)
\equiv
\frac{1}{\widetilde{\gamma}^i k_i +\widetilde{\gamma}^4 {k_4}_m}
= \frac{\widetilde{\gamma}^i\, k_i - \widetilde{\gamma}^4 \, {k_4}_m}{(\widetilde{\gamma}^i k_i)^2 - {k_4}_m^2}
= - \frac{\widetilde{\gamma}^i\, k_i - \widetilde{\gamma}^4 \, {k_4}_m}{(k_i)^2 + {k_4}_m^2} .
\end{equation}
The ultraviolet divergences of the anomalous terms in \eqref{eq:pimunu}
are regularized by the infinite set of Pauli--Villars-type fields $\xi_r(x)$, for $r\ne 0$, from \eqref{eq:Ireg-3D-gamma-matrices}.
The infrared divergences are regularized by imposing antiperiodic boundary conditions for the $\xi_r(x)$ fields ($r\in \mathbb{Z}$)
on the surface of a large ball $B^3$,
where the gauge fields $A_i(x)$ vanish
according to \eqref{eq:ball-condition-A}.

For a particular Fourier mode $n$ of the background gauge field,
the regularized two-point function is proportional to the
following expression:
\begin{eqnarray}
\hspace*{-10mm}&&
\pi^{ij} (p_n)\,\Big|^\text{(reg.)}
= \sum_{r=-\infty}^\infty (-1)^r \frac{1}{L}{\sum_{m=-\infty}^\infty} \int \frac{d^{3} k}{(2\pi)^3} \frac{\text{tr}\Big[\widetilde{\gamma}^i (\slashed k + M_r)\widetilde{\gamma}^j (\slashed k + \slashed p + M_r)
\Big]}{\Big(k_m^2 + M_r^2\Big)\Big((k_m+p_n)^2 + M_r^2\Big)},
\nonumber\\[1mm]
\hspace*{-10mm}&&
\label{eq:pimunu-reg}
\end{eqnarray}
with the short-hand notation
$\slashed p \equiv
\widetilde{\gamma}^i\, p_i - \widetilde{\gamma}^4\, {p_4}_n$
for the matrices \eqref{eq:3D-gamma-matrices},
which are Dirac gamma matrices in three spacetime dimensions
but not in four.
The factor $(-1)^r$ in (\ref{eq:pimunu-reg}) comes
from the Grassmann parity (\ref{Grassmann_Parity}) of the fields
and $M_r$ is given by \eqref{eq:regulator-masses-Mr}.
From now on, we drop the superscript `reg.'
as the regularization is manifest from having
\mbox{the sum over $r$.}

Introducing the Feynman parameter $x$ and changing
the momentum variable $k_\mu$ to $l_\mu$,
with $l_i \equiv k_i + x\, p_i$ and $l_4 \equiv k_4\,$,
we rewrite the expression for the vacuum-polarization kernel (\ref{eq:pimunu-reg}) as
\begin{eqnarray}
\label{pi-ij-pn}
\hspace*{-10mm}
\pi^{ij} (p_n) &=& \sum_{r=-\infty}^\infty (-1)^r \,
\int_0 ^1 dx \,
\frac{1}{L} \,\sum_{m=-\infty}^\infty \int \frac{d^{3} l}{(2\pi)^3}
\nonumber\\[1mm]
\hspace*{-10mm}
&&
\times\,
\text{tr}\Big[\widetilde{\gamma}^i\,
\Big(\widetilde{\gamma}^k\, l_k
- x\,\widetilde{\gamma}^k\, p_k
-\omega_m + M_r\Big)\, \widetilde{\gamma}^j\,
\nonumber\\[1mm]
\hspace*{-10mm}
&&
\cdot\,  \Big(\widetilde{\gamma}^k\,l_k
+ (1-x)\,\widetilde{\gamma}^k\,p_k
-\omega_m- \rho_n + M_r\Big)\Big]
\,
\Big(|\vec{l}|^2+ \Delta\Big)^{-2}\,,
\end{eqnarray}
with $p_n$, $\rho_n$, and $\rho_n^2$ from \eqref{eq:defs-pn-pn2-rhon}
and the further definitions
\begin{subequations}\label{pi-ij-pn-defs}
\begin{eqnarray}
l_m &\equiv& (\vec{l},\, \omega_m )\,,
\\[2mm]
\omega_m &\equiv& 2\pi m/L \,,
\\[2mm]
\Delta &\equiv& (\omega_m +x \rho_n )^2 + x(1-x)\,p_n^2  +M_r^2   \,.
\end{eqnarray}
\end{subequations}

The odd powers of the $l_i$ in the numerator of \eqref{pi-ij-pn}
vanish by symmetry reasons.
The term in \eqref{pi-ij-pn} with an odd number of $p_n$ momenta
in the numerator of the integrand is written as
\begin{eqnarray}
\widetilde{T}^{ij}(p_n) &=&
\sum_{r=-\infty}^\infty (-1)^r \frac{1}{L}{\sum_{m=-\infty}^\infty}
(-\omega_{m}+M_r) \int \frac{d^{3} l}{(2\pi)^3}  \int_0 ^1 dx
\nonumber\\[1mm]
&&
\times\,
\frac{\text{tr}[\widetilde{\gamma}^i \widetilde{\gamma}^j \widetilde{\gamma}^k]\,p_k
-  \text{tr}[\widetilde{\gamma}^i \widetilde{\gamma}^j \widetilde{\gamma}^4]~\rho _n}
{(|\vec{l}|^2+ \Delta)^2}\,. \label{tpn}
\end{eqnarray}
Part of the above equation still gives rise to
a finite $L$-independent term with an even number of $p_n$ momenta,%
\begin{eqnarray}
\hspace*{-10mm}
&&
\frac{1}{L}\,{\sum_{m=-\infty}^\infty} (-\omega_{m})
\int \frac{d^{3} l}{(2\pi)^3}  \int_0 ^1 dx \,
\sum_{r=-\infty}^\infty (-1)^r   ~
\frac{\text{tr}[\widetilde{\gamma}^i \widetilde{\gamma}^j \widetilde{\gamma}^k]\,p_k
-  \text{tr}[\widetilde{\gamma}^i \widetilde{\gamma}^j \widetilde{\gamma}^4]~\rho _n}
{(|\vec{l}|^2+ \Delta)^2}
\nonumber\\[1mm]
\hspace*{-10mm}
&&
\propto
\Big(\text{tr}[\widetilde{\gamma}^i \widetilde{\gamma}^j \widetilde{\gamma}^k]\, \rho_n\,p_k
-  \text{tr}[\widetilde{\gamma}^i \widetilde{\gamma}^j\widetilde{\gamma}^4]\,\rho_n \rho_n \Big)\,,
\label{eq:even-number-pn}
\end{eqnarray}
and we are left with the following term with an odd number of $p_n$ momenta:
\begin{equation}
\label{eq:T-of-pn-first1}
T^{ij}(p_n) =
\frac{1}{L}\,
{\sum_{m=-\infty}^\infty}\,
\int \frac{d^{3} l}{(2\pi)^3}  \int_0 ^1 dx
\,\sum_{r=-\infty}^\infty \,(-1)^r \,M_r
~ \frac{\text{tr}[\widetilde{\gamma}^i \widetilde{\gamma}^j \widetilde{\gamma}^k]\,p_k  - \text{tr}[\widetilde{\gamma}^i \widetilde{\gamma}^j]~\rho_n}{(|\vec{l}|^2+ \Delta)^2}\,,
\end{equation}
where we have taken care to move the $r$ sum inwards as it
must be performed first.

The $\rho_n$ term in the numerator of the integrand of (\ref{eq:T-of-pn-first1}) ultimately gives rise to a term
$\int_0^L d x^4 \int d^{3}x$ $\delta_{ij}\, A_i\, (\partial_4 A_j)$
in the effective gauge-field action, which is a total-derivative term
and vanishes due to the periodic boundary conditions
(\ref{eq:periodic-bcs}).
So, we are left with the following potentially CPT-violating term:
\begin{equation}
\label{eq:T-of-pn-first}
\hspace*{-5mm}
T_\text{anom}^{ij}(p_n) =
 \frac{1}{L}\,
{\sum_{m=-\infty}^\infty}\,
\int \frac{d^{3} l}{(2\pi)^3}  \int_0 ^1 dx
\,\sum_{r=-\infty}^\infty \,(-1)^r \,M_r~
\frac{\text{tr}[\widetilde{\gamma}^i \widetilde{\gamma}^j
\widetilde{\gamma}^k]\,p_k }{(|\vec{l}|^2+ \Delta)^2}.
\end{equation}

At this moment, we can
mention that the other regulator fields $\Psi_{s}$
from \eqref{eq:full-PV-type-lagrangian}
do not contribute to this potentially anomalous term
with an odd number of $p_n$ momenta, because the
trace of an odd number of Dirac matrices $\gamma^\mu$
vanishes. This is not the case for the trace of $\widetilde{\gamma}^i\,\widetilde{\gamma}^j\,\widetilde{\gamma}^k$,
as follows from relation \eqref{eq:3D-gamma-matrices-relation}.

We divide the sum over $m$ in  (\ref{eq:T-of-pn-first})
into two parts, namely, the sum over nonzero $m$ and the
single term
$m=0$ [this term is distinguished by having an infrared-divergent
momentum integral for the $r=0$ contribution, which is regularized
by antiperiodic boundary conditions as discussed a few lines below
\eqref{eq:propagator-S-of-km}].
The expression then reads
\begin{equation}
T_\text{anom}^{ij}(p_n) = T_0^{ij}(p_n) + T_\text{rest}^{ij}(p_n) \, ,
\label{Total-T-p}
\end{equation}
with
\begin{subequations}\label{eq:T-0-pn}
\begin{eqnarray}
T_0^{ij}(p_n) &=&
\frac{1}{L}\,
\int \frac{d^{3} l}{(2\pi)^3}  \int_0 ^1 dx
\,\sum_{r=-\infty}^\infty \,(-1)^r \,M_r~ \frac{\text{tr}[\widetilde{\gamma}^i \widetilde{\gamma}^j \widetilde{\gamma}^k]\,p_k }{(|\vec{l}|^2+ \Delta_0)^2} \, ,
\\[2mm]
\Delta_0 &\equiv& x \rho_n ^2 + x(1-x)\,p_n^2  +M_r^2 \,,
\end{eqnarray}
\end{subequations}
and
\begin{equation}
\label{eq:T-m-pn}
T_\text{rest}^{ij}(p_n) =
\frac{2}{L}\,
{\sum_{m=1}^\infty}\,
\int \frac{d^{3} l}{(2\pi)^3}  \int_0 ^1 dx
 \,\sum_{r=-\infty}^\infty \,(-1)^r \,M_r~ \frac{\text{tr}[\widetilde{\gamma}^i \widetilde{\gamma}^j \widetilde{\gamma}^k]\,p_k }{(|\vec{l}|^2+ \Delta)^2} \,,
\end{equation}

First, consider the $m=0$ contribution \eqref{eq:T-0-pn}.
In order to compute the sum over $r$,
we use the following representation
(defining $l \equiv |\vec{l}|$):
\begin{subequations}
\begin{eqnarray}\label{eq:S0}
S_{0} &=& \sum_{r=-\infty}^\infty
\frac{(-1)^r M_r}
{\Big( |\vec{l}|^2 + (x \rho_n) ^2 + x(1-x)\,p_n^2  +M_r^2 \Big)^2}
\nonumber\\
&=&
 - \frac{1}{2\,l}
\frac{\partial}{\partial l}\sum_{r=-\infty}^\infty
\frac{(-1)^r M_r}
{\Big( l^2 + (x \rho_n) ^2 + x(1-x)\,p_n^2  +M_r^2\Big)}
\nonumber\\
&=&
 - \frac{1}{2\,l}\,\frac{M}{M^2}
\frac{\partial}{\partial l}\sum_{r=-\infty}^\infty
\frac{(-1)^r \, r^2}
{\left( \tau^2  + r^4\right)}\,,
\end{eqnarray}
with
\begin{eqnarray}
\tau^2 &\equiv& \left[l^2+ (x \rho_n) ^2 + x\,(1-x)\,p_n^2 \right]/M^2
\equiv l^2/M^2 +\kappa   \,,
\end{eqnarray}
\end{subequations}
and the following result (for $\tau\ne 0$):
\begin{subequations}\label{eq:r-sum-tau2-f-of-tau}
\begin{eqnarray}
\label{eq:r-sum-tau2}
\hspace*{-0mm}
\sum_{r=-\infty}^\infty \frac{(-1)^r \, r^2}{\tau^2+r^4}
&=&
f(\tau) \,,
\\[2mm]
\label{eq:f-of-tau}
\hspace*{-0mm}
f(\tau) &\equiv&
\frac{\pi}{2 \,\sqrt{\tau}}\,
\Bigg(\frac{\exp(i\pi/4)}
{\text{sinh}\big[\exp(-i\pi/4)\,\pi \, \sqrt{\tau} \,\big]}
+ \text{c.c.} \Bigg).
\end{eqnarray}
\end{subequations}

Remark that the first sum in \eqref{eq:S0}
contains an extra factor $M_r$ in the numerator compared to
Eq.~(11) of Ref.~\cite{FrolovSlavnov1993} and this is the reason
for demanding the $r^2$ behavior in the regulator masses
$M_r$ in \eqref{eq:regulator-masses-Mr}. We then find the same
type of $1/\text{sinh}$ behavior in \eqref{eq:f-of-tau}
as in Eq.~(14) of Ref.~\cite{FrolovSlavnov1993}, which,
in both cases, provides an exponential cutoff of the momentum
integrals.

With result \eqref{eq:r-sum-tau2-f-of-tau},
expression (\ref{eq:T-0-pn}) reduces to
\begin{eqnarray}
\hspace*{-8mm}
T_0^{ij}(p_n) &=&
- \frac{1}{4 \pi^2 L}\,\frac{M}{M^2}
\int_0 ^1 dx \int_0 ^\infty l d l
~ \frac{\partial}{\partial l}
\Big[ f(\tau)\Big]\,
\text{tr}[\widetilde{\gamma}^i\, \widetilde{\gamma}^j\, \widetilde{\gamma}^k]\,p_k
\nonumber\\[1mm]
\hspace*{-8mm}
&=& - \frac{1}{4 \pi^2 L} \,\frac{M}{|M|}
\int_0 ^1 dx \int_0 ^\infty d~ \eta ~ \eta
~ \frac{\partial}{\partial \eta}
\Big[ f(\tau)\Big]\,
\text{tr}[\widetilde{\gamma}^i\, \widetilde{\gamma}^j\, \widetilde{\gamma}^k]\,p_k\,,
\label{aboveequation}
\end{eqnarray}
in terms of the dimensionless variable $\eta \equiv l/|M|$.
In the following, we assume positive $M$
(the related ambiguity in the anomalous term, here by a
factor $M/|M|$, is discussed
further in the first paragraph of Sec.~\ref{sec:Conclusion}).

In the regularization procedure, we consider the regulator mass scale $M$
to be much larger than a typical momentum component of the gauge field,
$M^2 \gg p_n^2$, so that we can take
$\kappa \equiv \left[ (x\rho_n)^2 + x\,(1-x)\, p_n^2\right]/M^2 \to 0^{+}$
in the rest of the calculation and the $x$ integral
in (\ref{aboveequation}) becomes trivial. Using
\begin{equation}\label{eq:trace-gammatilde-ijk}
\text{tr}[\widetilde{\gamma}^i\, \widetilde{\gamma}^j\, \widetilde{\gamma}^k] = 2\,   \epsilon^{ijk}\,,
\end{equation}
we then rewrite (\ref{aboveequation}) for positive $M$ as
\begin{eqnarray}\label{eq:T-of-pn}
T_0^{ij}(p_n) &=& - \frac{1}{2 \pi^2 L}
\left(\int_0 ^\infty d~ \eta ~ \eta
~ \frac{\partial}{\partial \eta} \Big[ f(\eta)\Big]\right) \,
\epsilon^{ijk}\,p_k\,.
\end{eqnarray}
The $\eta$ integral in \eqref{eq:T-of-pn} gives a factor $\pi /2$
and the final result for the $m=0$ sector reads
\begin{equation}
T_0^{ij}(p_n)= -\frac{1}{4 \pi L} ~ \epsilon^{ijk}\,p_k\,.
\label{eq:T-pn-final1}
\end{equation}

Now turn to the $m\neq 0$ sum (\ref{eq:T-m-pn}),
\begin{subequations}
\begin{equation}
\label{eq:T-m-pn1}
T_\text{rest}^{ij}(p_n) = \frac{1}{L}\,
{\sum_{m\neq 0}}\,
\int \frac{d^{3} \eta}{(2\pi)^3}  \int_0 ^1 dx
\,\sum_{r=-\infty}^\infty \,(-1)^r \,r^2~ \frac{\text{tr}[\widetilde{\gamma}^i \widetilde{\gamma}^j \widetilde{\gamma}^k]\,p_k }{\Big(|\vec{\eta}|^2+ \Delta_M\Big)^2} \,,
\end{equation}
with
\begin{equation}
|\vec{\eta}|^2 \equiv |\vec{l}|^2/M^2 \,.
\end{equation}
and
\begin{equation}
\Delta_M \equiv \Big[(\omega_m + x \rho_n) ^2 + x(1-x)\,p_n^2\Big]/M^2
+ r^4 \sim \omega_m ^2 /M^2  + r^4 \,,
\end{equation}
\end{subequations}
for $p_n^2/M^2 \to 0$.
With large $M$, we can treat $\omega_m / M \equiv l_4$ as a continuous variable and rewrite \eqref{eq:T-m-pn1} as follows:
\begin{eqnarray}
\label{eq:T-m-pn2}
T_\text{rest}^{ij}(p_n) &=& \frac{M}{2\pi}\,
\int dl_4 ~ \int \frac{d^{3} \eta}{(2\pi)^3}  \int_0 ^1 dx
\,\sum_{r=-\infty}^\infty \,(-1)^r \,r^2~ \frac{\text{tr}[\widetilde{\gamma}^i \widetilde{\gamma}^j \widetilde{\gamma}^k]\,p_k }{(\lambda^2 + r^4)^2}
\nonumber\\[1mm]
&=&  M \,
\int \frac{d^{4} \lambda}{(2\pi)^4}  \int_0 ^1 dx
\,\sum_{r=-\infty}^\infty \,(-1)^r \,r^2~ \frac{\text{tr}[\widetilde{\gamma}^i \widetilde{\gamma}^j \widetilde{\gamma}^k]\,p_k }{(\lambda^2 + r^4)^2} \,,
\end{eqnarray}
in terms of the dimensionless variable
$\lambda^2 \equiv |\vec{\eta}|^2 + (l_4)^2$.

In order to compute the sum over $r$ in \eqref{eq:T-m-pn2},
we again use the following representation:
\begin{eqnarray}
\sum_{r=-\infty}^\infty
\frac{(-1)^r \, r^2}
{(\lambda^2 + r^4)^2}
&=&
 - \frac{1}{2\lambda}
\frac{\partial}{\partial\lambda}\sum_{r=-\infty}^\infty
\frac{(-1)^r \, r^2}
{(\lambda^2 + r^4)}\,,
\end{eqnarray}
where the last sum has the same form as \eqref{eq:r-sum-tau2} and
equals $f(\lambda)$ in terms of the function $f$
defined by (\ref{eq:f-of-tau}).
As mentioned above, the $x$ integral in expression
(\ref{eq:T-m-pn2}) is trivial and the expression reduces to
\begin{eqnarray}\label{eq:T-pn-lambda-integral}
T_\text{rest}^{ij}(p_n) &=& - \frac{M}{16 \pi^2}
\left(\int_0 ^\infty d \lambda~ \lambda^2 ~
~ \frac{\partial}{\partial \lambda}
\Big[ f(\lambda)\Big]\right)\,
\text{tr}[\widetilde{\gamma}^i\, \widetilde{\gamma}^j\, \widetilde{\gamma}^k]\,p_k
\nonumber\\[1mm]
&=&- \frac{M}{8 \pi^2}
\left(\int_0 ^\infty  d \lambda ~ \lambda^2 ~
~ \frac{\partial}{\partial \lambda}
\Big[ f(\lambda)\Big]\right)\,
\epsilon^{ijk}\,p_k \,,
\label{aboveequation2}
\end{eqnarray}
where the last step uses \eqref{eq:trace-gammatilde-ijk}.
The $\lambda$
integral in \eqref{eq:T-pn-lambda-integral} gives the following  factor:
\begin{equation}
\xi =14\,\zeta (3)/\pi ^2 \approx 1.70511\,,
\label{eq:xi}
\end{equation}
and the final expression reads
\begin{equation}
T_\text{rest}^{ij}(p_n)=
-\xi\, M\, \frac{1}{ 8 \pi^2 } ~ \epsilon^{ijk}\,p_k\,.
\label{eq:T-pn-final2}
\end{equation}

Combining (\ref{eq:T-pn-final1}) and (\ref{eq:T-pn-final2})
gives the end result for the anomalous
vacuum-polarization kernel \eqref{Total-T-p},
\begin{equation}
T_\text{anom}^{ij}(p_n)=
-\frac{1}{4 \pi L} ~ \epsilon^{ijk}\,p_k\,
-\xi\, M\, \frac{1}{ 8 \pi^2 } ~ \epsilon^{ijk}\,p_k\,,
\label{eq:T-pn-final3}
\end{equation}
with the constant $\xi$ given by \eqref{eq:xi}
and the regulator mass scale $M$ entering the
Pauli--Villars-type masses \eqref{eq:regulator-masses-Mr}.
The first term in (\ref{eq:T-pn-final3})
is $L$-dependent and finite,
whereas the second term is $L$-independent and divergent
as the regulator mass scale $M$ is taken to infinity.
As regards the $M$-dependence of this second term,
note that, for four-dimensional quantum electrodynamics,
the vacuum polarization from the standard Pauli-Villars
regularization also has an $M$-dependent contribution;
cf. Eq.~(A.6) in Ref.~\cite{FrolovSlavnov1993}.
A suitable renormalization procedure
is to subtract the same result at a reference value
$L_\text{ref}$ and to take $L_\text{ref}\to\infty$
corresponding to Minkowski spacetime
(cf. Sec.~4.2 of Ref.~\cite{BirrellDavies1982}).
This renormalization procedure then eliminates
the second term in (\ref{eq:T-pn-final3})
and we are left with the first term only,
\begin{equation}
T_\text{anom}^{ij}(p_n)\,\Big|^\text{(renorm.)}
= -\frac{1}{4 \pi L} ~ \epsilon^{ijk}\,p_k\,.
\label{eq:T-pn-final}
\end{equation}

Now replace the single left-handed fermion $\psi_L$
by the 48 left-handed fermions of the chiral $U(1)$ gauge theory \eqref{eq:U1theory-G-RL}, with the same regularization
for each of these 48 fermions. Using (\ref{eq:T-pn-final}),
we then obtain the following local expression for the
effective gauge-field action \eqref{factor1} to order $e^2$:
\begin{eqnarray}
\mathcal{T}_\text{anom}^\text{\,(renorm.)} =  i \,F\,e^2\,
\frac{1}{8\pi L} \int_0^L dx^4 \int_{\mathbb{R}^3} d^{3} x~
\epsilon^{ijk}\,A_i(x) ~ \partial_j ~ A_k(x)\,,
\label{eq:mathcalT-anom}
\end{eqnarray}
with an overall numerical factor $F$ from \eqref{eq:sum-charges-square-F}
due to the contributions of
all chiral fermions of the theory \eqref{eq:U1theory-G-RL}.
The result \eqref{eq:mathcalT-anom} gets a further
factor $i$ for spacetime metrics with Lorentzian signature and
a spatial coordinate $x^4\in S^1$ (see also the discussion of the
last paragraph in Sec.~\ref{sec:Conclusion}).
The local effective-action term \eqref{eq:mathcalT-anom} is
the main result of the perturbative calculation.

For gauge fields $A_\mu(x)$ of local support,
the term \eqref{eq:mathcalT-anom} is invariant under local
Abelian gauge transformations,
\begin{eqnarray}
A_\mu(x) \to A_\mu(x) + i \,\partial_\mu\,\zeta(x)\,,
\label{eq:Abelian-gauge-transformation}
\end{eqnarray}
with arbitrary real gauge parameters $\zeta(x)$ that are
$x^4$-periodic, $\zeta(\vec{x},\,0)=\zeta(\vec{x},\,L)$.
As mentioned in Sec.~\ref{sec:Setup-problem},
the Abelian
holonomy \eqref{eq:trivial-holonomy-A} is gauge-invariant
under these periodic transformations.
The perturbative calculation of this subsection can, in principle,
be extended to the non-Abelian theory \eqref{eq:SO10theory-G-RL}
and we expect a further cubic term in addition to the quadratic
term of \eqref{eq:mathcalT-anom}, in order to maintain invariance
under ``small'' gauge transformations
(see Sec.~4 in Ref.~\cite{Klinkhamer1999} for further discussion).

%%\newpage%%tmp
\subsection{Lorentz and CPT violation}
\label{subsec:Lorentz and CPT violation}

For arbitrary gauge fields $A_\mu (x)$
with trivial holonomies \eqref{eq:trivial-holonomy-A}
in the chiral $U(1)$ gauge theory \eqref{eq:U1theory-G-RL}
with a Lorentzian metric signature,
our result \eqref{eq:mathcalT-anom} gives the following term
in the effective gauge-field action at the one-loop level:
\begin{subequations}\label{eq:Gamma-anom-perturbative-Gamma-CS-like}
\begin{eqnarray}
\label{eq:Gamma-anom-perturbative}
\Gamma_\text{anom}[A]
&=&
-2\pi\,F\,e^2\,\Gamma_\text{CS-like}[A]\,,
\\[2mm]
\label{eq:Gamma-CS-like}
\Gamma_\text{CS-like}[A]
&\equiv&
\frac{1}{L}\int_0^L dx^4 \int_{\mathbb{R}^3} d^{3}x
~ \omega_\text{CS}[A(\vec{x}, x^4)]\,,
\end{eqnarray}
\end{subequations}
in terms of the Chern--Simons density~\cite{ChernSimons1974}
\begin{equation}
\label{eq:omega-CS}
\omega_\text{CS}[A(\vec{x}, x^4)]
\equiv
\frac{1}{16\pi^2} \,
\epsilon^{ijk}\,A_i(\vec{x}, x^4)\,\partial_j\, A_k(\vec{x}, x^4)\,.
\end{equation}
The numerical factor $F$ in \eqref{eq:Gamma-anom-perturbative}
is given by \eqref{eq:sum-charges-square-F}.

A topological Chern--Simons
term $\Omega_\text{CS}=\int \omega_\text{CS}$
is defined only for an odd number of spacetime dimensions~\cite{ChernSimons1974}.
The action term \eqref{eq:Gamma-anom-perturbative-Gamma-CS-like}
holds, however, in four spacetime dimensions. Hence, the
qualification ``Chern--Simons-like'' (abbreviated as ``CS-like'')
used in \eqref{eq:Gamma-CS-like} and elsewhere.
The action term \eqref{eq:Gamma-anom-perturbative-Gamma-CS-like} is nontopological in the sense that it has a nontrivial dependence
on the spacetime metric or vierbein (see Sec.~6.6 of Ref.~\cite{Klinkhamer2005} for further discussion and references).

Observe that the integrand of (\ref{eq:Gamma-CS-like}) is
proportional to
$\epsilon^{\,\mu\nu\rho 4}\,A_{\mu}(x)\,\partial_{\nu}\, A_{\rho}(x)$,
which has the spacetime index `$4$' singled-out.
This term is, therefore, Lorentz noninvariant.
Next, recall that the CPT transformation of an anti-Hermitian
gauge field is given
by~\cite{Klinkhamer1999}
\begin{equation}
A_\mu (x)\rightarrow A_\mu (-x)\,. \label{cptaH}
\end{equation}
The term (\ref{eq:Gamma-CS-like})
changes sign under a CPT transformation (\ref{cptaH}).
The Lorentz-violating term (\ref{eq:Gamma-CS-like}) is,
therefore, also CPT-odd [the Lorentz-invariant
Maxwell term $(\partial_\mu A_\nu -\partial_\nu A_\mu)\,
(\partial^\mu A^\nu -\partial^\nu A^\mu)$ is CPT-even].

%%\newpage%%tmp
\section{Nonperturbative approach}
\label{sec:Nonperturbative-approach}

\subsection{Lattice setup}
\label{subsec:Lattice-setup}

In our calculation, we consider a chiral gauge theory which is defined over a four-dimensional spacetime manifold $M = \mathbb{R}^3 \times S^1$, with noncompact coordinates
$x^1,\, x^2,\, x^3 \in \mathbb{R}$ and compact coordinate $x^4 \in [0 , L]$.
Initially, the metric is taken to be the Euclidean flat metric
$g_{\mu\nu}=[\text{diag}(1,\, 1,\, 1,\, 1)]_{\mu\nu}$.
The vierbeins (tetrads) are trivial and given by
\begin{equation}
e^a_\mu (x)= \delta^a_\mu\,,
\end{equation}
with the Lorentz index $a=1,\, 2,\, 3,\, 4$ and the
Einstein index $\mu = 1,\, 2,\, 3,\, 4$.

We consider, in particular, chiral gauge theories that are free
of gauge anomalies.
As mentioned in Sec.~\ref{sec:Setup-problem},
we can take the $SO(10)$ chiral gauge theory \eqref{eq:SO10theory-G-RL}.
But, in order to be sure of
having a well-defined lattice gauge theory~\cite{Luscher1998b},
we restrict ourselves to the Abelian $U(1)$
theory  \eqref{eq:U1theory-G-RL}.
The actual calculation in the rest of this section is performed
for a single left-handed fermion $\psi_L$ with unit $U(1)$ charge, $q=e$.
Only the final result \eqref{eq:Gamma-CS-like-nonperturbative}
is extended to all chiral fermions of
the theory \eqref{eq:U1theory-G-RL}.

To regularize the ultraviolet divergences of this gauge theory,
a rectangular hypercubic lattice with lattice spacing $a$ is introduced,
\begin{subequations}
\begin{equation}\label{eq:4D-hypercubic-lattice}
(x^1,\, x^2,\, x^3,\, x^4)\equiv (\vec{x}, x^4)
= (\vec{n}\, a, n_4\,  a),
\end{equation}
with integers
\begin{equation}
n_1,\,  n_2,\,  n_3 \in [0,\, N^\prime]\,,\quad n_4 \in [0,\,  N]\,.
\end{equation}
\end{subequations}
The fermion fields and link variables are periodic with respect to
the $x^4$ coordinate,
\begin{subequations}\label{eq:psi-pbcspsibar-pbcs-U-pbcs}
\begin{eqnarray}
\psi(x^1,\, x^2,\, x^3,\, L)&=&
\psi(x^1,\, x^2,\, x^3,\,0)   \,,
\\[2mm]
\overline{\psi}(x^1,\, x^2,\, x^3,\, L) &=&
\overline{\psi}(x^1,\, x^2,\, x^3,\,0) \,,
\\[2mm]
U_\mu(x^1,\, x^2,\, x^3,\, L) &=&
U_\mu(x^1,\, x^2,\, x^3,\,0)  \,,
\end{eqnarray}
\end{subequations}
with $L \equiv N\,a$.
For the other coordinates, the link variables are again periodic but the fermion fields are taken to be antiperiodic, for example,
\begin{subequations}\label{eq:psi-apbcspsibar-apbcs-U-pbcs}
\begin{eqnarray}
\psi(L^\prime,\, x^2,\, x^3,\, x^4)&=&-\psi(0,\, x^2,\, x^3,\, x^4)   \,,
\\[2mm]
\overline{\psi}(L^\prime,\, x^2,\, x^3,\, x^4) &=&-\overline{\psi}(0,\, x^2,\, x^3,\, x^4)   \,,
\\[2mm]
U_\mu(L^\prime, x^2, x^3, x^4) &=& U_\mu(0,\, x^2,\, x^3,\, x^4)   \,,
\end{eqnarray}
\end{subequations}
and similarly for the other coordinates $ x^2$ and $x^3$.

The assumptions \eqref{eq:zero-A4-condition}
for the continuum gauge fields translate into the following
conditions on the link variables of the lattice:
\begin{subequations}\label{eq:condition-links}
\begin{eqnarray}
U_i(x) &=& U_i(x^1,x^2, x^3, x^4)\,,\;\;\text{for}\;\;i=1,\,2,\,3 \,,
\\[2mm]
U_4 (x) &=& \id\,.
\end{eqnarray}
\end{subequations}
As mentioned before,
such link variables can be obtained by a gauge transformation only
if there are trivial holonomies,
\begin{equation}\label{eq:trivial-holonomy-U}
H_4(x^1,\,x^2,\, x^3)
\equiv \prod_\text{links} \,U_4(x^1,\,x^2,\, x^3,\, x^4)
= \id \,,
\end{equation}
where the product runs over all $U_4$ links in the 4-direction
at a fixed value of $\vec{x}$
(for non-Abelian gauge groups,
these non-commuting matrices $U_4$ are ordered along the path).

The anti-Hermitian Abelian gauge field $A_\mu$
of the continuum
and the $U(1)$ link variable $U_\mu$ of the lattice
are related as follows~\cite{MontvayMunster1997}:%
\begin{equation}\label{eq:Umu-Amu}
U_\mu (x)
= \exp \left[ e\,\int_x^{x+a\,\widehat{\mu}} dy\, A_\mu(y) \right]
\approx \exp \Big[e\,a\,A_\mu (x+a\,\widehat{\mu}/2)\Big]\,,
\end{equation}
where the integration variable $y$ in the second expression
runs over a straight line between the spacetime points $x$
and $x+a\,\widehat{\mu}$,
with unit vector $\widehat{\mu}$ in the $\mu$ direction.
In \eqref{eq:Umu-Amu},
$e$ is the dimensionless electric charge of the fermion.

Recall from Sec.~\ref{sec:Setup-problem} that
Latin spacetime indices $i,\,j,\,k,\, l$, etc. run
over the coordinate labels $1,\, 2,\, 3$,
and Greek spacetime indices $\mu,\, \nu,\, \rho$, etc.
over the labels $1,\, 2,\, 3,\, 4$,
and that we use natural units with $\hbar = c= 1$.

%%\newpage%%tmp
\subsection{Chiral fermions on the lattice}
\label{subsec:Chiral fermions on the lattice}

\subsubsection{Ginsparg--Wilson relation}
\label{subsubsec:Ginsparg--Wilson relation}

In order to avoid the fermion-doubling problem,
Wilson introduced an operator,
now known as the Wilson--Dirac operator~\cite{MontvayMunster1997},
which includes a term of second order in the difference operators,
\begin{equation}\label{eq:wilson-dirac-operator}
D_W = \frac{1}{2} \sum_{\mu=1}^{4}
\Big[\gamma_\mu(\nabla_\mu +\nabla_\mu^\ast)
+ s\,a\nabla_\mu \nabla_\mu^\ast \Big]\, ,
\end{equation}
with $4\times 4$ Dirac matrices $\gamma_\mu$ and
a parameter $s$ to be described below. Here,
the gauge-covariant derivatives of the continuum are replaced by gauge-covariant forward and backward difference operators on the lattice,
\begin{subequations}\label{eq:}
\begin{eqnarray}
\nabla_\mu \psi (x) &\equiv& \frac{1}{a}
\Big( R [U_\mu (x)] \psi (x+ a\,\widehat{\mu})- \psi(x) \Big)  \,,
\\[2mm]
\nabla_\mu^\ast \psi (x) &\equiv& \frac{1}{a}
\Big( \psi(x)- R [U_\mu (x- a\,\widehat{\mu})]^{-1} \psi (x- a\,\widehat{\mu})  \Big)   \,,
\end{eqnarray}
\end{subequations}
where $R$ is a unitary representation of the gauge group.

The Wilson parameter $s$ in (\ref{eq:wilson-dirac-operator})
takes the values $s=\pm 1$. For definiteness, we choose
\begin{equation}
s=-1. \label{wilson parameter}
\end{equation}
The $s$ term in \eqref{eq:wilson-dirac-operator} breaks, however, the chiral invariance of the theory.
In order to restore the chiral symmetry, Ginsparg and Wilson
suggested to implement the following relation \cite{GinspargWilson1982}:
\begin{equation}
D\,\gamma_5+\gamma_5\, D = a\, D\, \gamma_5\, D \,,
\label{Ginsparg-Wilsonrelation}
\end{equation}
which is known as the Ginsparg--Wilson relation.

Sixteen years after Ginsparg and Wilson proposed relation
\eqref{Ginsparg-Wilsonrelation}, Neuberger explicitly constructed a corresponding
operator~\cite{Neuberger1998a, Neuberger1998b},
\begin{equation}\label{eq:DofU-Neuberger}
D[U]= \frac{1}{a}\Big(\id - V[U]\Big)\,,
\end{equation}
in terms of an appropriate unitary operator $V$.
Apart from satisfying the Ginsparg--Wilson relation
\eqref{Ginsparg-Wilsonrelation}, the operator $V$ should also be $\gamma_5$-Hermitian,
\begin{equation}
V^\dagger = \gamma_5\, V\, \gamma_5\,.
\end{equation}
In terms of the Wilson--Dirac operator $D_W$ from
\eqref{eq:wilson-dirac-operator}, this operator $V$ reads%
\begin{subequations}\label{eq:def-V-X}
\begin{eqnarray}
\label{eq:def-V}
V &=&  X\, (X^\dagger X)^{-1/2}=
\int_{-\infty}^{\infty} \frac{dt}{\pi}\, \Big(t^2 + X^\dagger X\Big)^{-1}   \,,
\\[2mm]
\label{eq:def-X}
X  &\equiv&\id - a\,D_W   \,.
\end{eqnarray}
\end{subequations}

\subsubsection{Lattice fermion action}
\label{subsubsec:Lattice fermion action}

The lattice fermion action with a Ginsparg--Wilson operator $D[U]$
defined by \eqref{eq:DofU-Neuberger} and \eqref{eq:def-V-X},
\begin{equation}
S_F[\overline{\psi} , \psi , U ] =  a^4 \sum_x  ~\overline{\psi}(x) D[U] \psi(x)]\,, \label{fermion action}
\end{equation}
is invariant under the following infinitesimal transformations \cite{Luscher1998a}:
\begin{subequations}\label{eq:psi-psibar-Luscher-transf}
\begin{eqnarray}
\psi(x) &\rightarrow& \psi(x)+\delta\psi(x)   \,,
\\[2mm]
\overline{\psi}(x)&\rightarrow& \overline{\psi}(x)+\delta\overline{\psi}(x)   \,,
\end{eqnarray}
\end{subequations}
with
\begin{subequations}\label{eq:deltapsi-deltapsibar-Luscher-transf}
\begin{eqnarray}
\label{eq:deltapsi-deltapsibar-Luscher-transf-delta-psi}
\delta\psi(x) &=&
i \varepsilon\,  \gamma_5 V \, \psi(x)
\equiv i \varepsilon\,  \widehat{\gamma}_5 \, \psi(x)   \,,
\\[2mm]
\label{eq:deltapsi-deltapsibar-Luscher-transf-delta-bar-psi}
\delta\overline{\psi}(x) &=&
i \varepsilon\, \overline{\psi}(x) \gamma_5 \,,
\end{eqnarray}
\end{subequations}
where $\varepsilon$ is an infinitesimal parameter.
The operator $\widehat{\gamma}_5 $, as defined in
\eqref{eq:deltapsi-deltapsibar-Luscher-transf-delta-psi},
is a Hermitian unitary operator with eigenvalues $\pm 1$.

A chiral gauge theory for left-handed fermions on the lattice
can be constructed by imposing the following
constraints~\cite{Luscher1998b}:
\begin{subequations}\label{eq:fermion-constraints-Luscher}
\begin{eqnarray}
 \psi(x)= \widehat{P}_{-}\, \psi(x)  \,,
\\[2mm]
\overline{\psi}(x) = \overline{\psi}(x)\, P_{+}   \,,
\end{eqnarray}
\end{subequations}
with the projection operators
\begin{subequations}\label{eq:projection-operator-Luscher}
\begin{eqnarray}\label{eq:projection-operator-Luscher-Phat}
\widehat{P}_\pm \equiv \frac{1}{2}\,(1\pm \widehat{\gamma_5})   \,,
\\[2mm]
P_\pm \equiv  \frac{1}{2}\,(1\pm \gamma_5)   \,,
\end{eqnarray}
\end{subequations}
where $\widehat{\gamma}_5 $ has been defined in
\eqref{eq:deltapsi-deltapsibar-Luscher-transf-delta-psi}.

\subsubsection{Discrete transformations}
\label{subsubsec: Discrete transformations}

On the hypercubic spacetime lattice,
there are certain symmetry transformations. Specifically,
these lattice symmetries are
\begin{enumerate}
\item[(i)]
the translations by an integer multiple of the lattice spacing $a$ in the direction of one of the four coordinate axes,
\item[(ii)]
the rotations by an integer multiple of the angle $\pi/2$ in hyperplanes spanned by two axes,
\item[(iii)]
the parity transformation,
\item[(iv)]
the time-reversal transformation,
\item[(v)]
the charge-conjugation  transformation.
\end{enumerate}

We now give the parity, time-reversal, and charge-conjugation
transformations for the link variable, considering the $x^1$ coordinate to
be the time coordinate for the Lorentzian metric signature
and using the notation
$x=(x^1,\,x^2,\, x^3,\, x^4) \equiv  (x^1,\, \widetilde{x})$.
The parity-transformed link variable is
\begin{subequations}
\begin{equation}
{U_\mu}^{\mathcal{P}}(x^1,\, \widetilde{x})=
\left\{
\begin{array}{ll}
 U_\mu^\dagger (x^1,\,  -\widetilde{x}-a\,\widehat{\mu})\,,&
\;\;\text{for}\;\;\mu = 2,\, 3,\,4\,, \\[1mm]
U_\mu(x^1,\,  -\widetilde{x})\,, & \;\;\text{for}\;\;\mu =1\,,
\end{array}
\right.
\end{equation}
the time-reflected link variable is
\begin{equation}
{U_\mu}^{\mathcal{T}}(x^1,\, \widetilde{x})=
\left\{
\begin{array}{ll}
U_\mu^\ast (-x^1,\,  \widetilde{x})\,, &
\;\;\text{for}\;\;\mu = 2,\, 3,\,4\,, \\[1mm]
U^t_\mu(-x^1-a,\,  \widetilde{x})\,, & \;\;\text{for}\;\;\mu =1\,,
\end{array}
\right.
\end{equation}
and the charge-conjugated link variable is
\begin{equation}
{U_\mu}^{\mathcal{C}}(x^1,\, \widetilde{x})=  U_\mu^\ast(x^1,\, \widetilde{x}).
\end{equation}
\end{subequations}
Hence, the combined CPT transformation on a link variable is given by
\begin{equation}
{U_\mu}^\theta (x)=  U_\mu^\dagger (-x-a\,\widehat{\mu}).
\end{equation}

\subsubsection{Integration measure}
\label{subsubsec:Integration measure}

The fermionic integration measure is the product of all integration measures at the sites of the hypercubic lattice,
\begin{equation}
\mathcal{D}\psi(x) = \prod_{x,\alpha} d\psi_\alpha(x)\,,
\quad
\mathcal{D} \overline{\psi}(x) = \prod_{x,\alpha}  d\overline{\psi}_\alpha(x)\,,
\end{equation}
with a multi-index $\alpha$ containing the
spinor, gauge, and flavor indices.

The fermionic fields can be expanded as follows:
\begin{equation}
\psi(x)= \sum_j v_j(x)\, c_j, ~~
\overline{\psi}(x)= \sum_k \bar{c}_k\, \bar{v}_k(x)\,,
\end{equation}
where the  $c_j$ and $\bar{c}_k$ are Grassmann-valued coefficients and
the $v_j(x)$ and $\bar{v}_k(x)$ are two orthonormal bases of
complex-valued spinorial functions. The integration measure is
then given by
\begin{equation}
\mathcal{D}\psi(x) = \prod_j dc_j\,,
\quad
\mathcal{D} \overline{\psi}(x) = \prod_k d\bar{c}_k\,. \label{measure}
\end{equation}

But this integration measure is not unique. Let $\mathcal{U}$ be a unitary operator which diagonalizes the operator $\widehat{\gamma}_5$,
\begin{equation}
\mathcal{U}^\dagger\,  \widehat{\gamma}_5\, \mathcal{U}
= \gamma_5\,,
\end{equation}
where $\gamma_5$ on the right-hand side is diagonal in the Weyl
representation of the Dirac gamma matrices. Then,
the basis spinors $v_j$ are
\begin{equation}
v_j(x)= \mathcal{U}\, \chi_j(x)\,,
\end{equation}
where the $\chi_j$ form a complete canonical spinor basis and
satisfy the chirality constraint
\begin{equation}
\widehat{P}_{-}\, \chi_j(x)= \chi_j(x)\,.
\end{equation}
Now,
$\mathcal{U}^\prime=\mathcal{U}Q$ is also
a diagonalization operator if $Q$ has the following form:
\begin{equation}
Q= \left( \begin{matrix}
  Q_1 & 0  \\
  0 & Q_2  \\
 \end{matrix} \right), ~~
Q_1^\dagger\, Q_1 = \id, ~~ Q_2^\dagger\, Q_2= \id\,,
\end{equation}
where
$Q_1$ and $Q_2$ are $2\times2$ block matrices in spinor space.
If the basis vectors change as
\begin{subequations}\label{eq:vj-change}
\begin{eqnarray}
v^\prime_j(x)= \sum_i v_i(x) \,\mathcal{Q}_{ij}  \,,
\end{eqnarray}
with
\begin{eqnarray}
\mathcal{Q}_{ij} \equiv
a^4 \sum \chi_i^\dagger (x)\, Q\, \chi_j(x)   \,,
\end{eqnarray}
\end{subequations}
then the measure (\ref{measure}) changes by a factor $\det \mathcal{Q}$, which is a phase factor since $\mathcal{Q}$ is unitary.

%%\newpage%%tmp
\subsection{Effective action and CPT transformation}
\label{subsec:Effective action and CPT transformation}

\subsubsection{Effective action}
\label{subsubsec:Effective action}

As in Sec.~\ref{sec:Perturbative-approach}, we calculate
the effective gauge-field action by integrating out the chiral fermions, while maintaining gauge invariance.
In lattice gauge theory, the Euclidean path integral is given by:
\begin{equation}
\exp(-\Gamma[ U ]) = \frac{1}{Z}\int \prod_x \mathcal{D} \overline{\psi}(x)\prod_x \mathcal{D} \psi(x) ~ \exp\left(-S_F [\overline{\psi}, \psi, U] \right), \label{effective field action for lattice}
\end{equation}
where $S_F$ is defined by (\ref{fermion action}). The normalization constant $Z$ ensures that $\Gamma[ \id ] =0$ for the constant-link-variable configuration $U_\mu(x)=\id$.

We Fourier expand the chiral fermionic fields as follows:
\begin{subequations}\label{eq:psi-psibar-Fourier-expansions}
\begin{eqnarray}
\psi(x) &=&
\frac{1}{L}\sum_n \psi_n(x^1,x^2,x^3) \, e^{2\pi i n x^4/L}\,,
\\[2mm]
\overline{\psi}(x) &=&
\frac{1}{L}\sum_n \overline{\psi}_n(x^1,x^2,x^3)\,  e^{-2\pi i n x^4/L} \,,
\end{eqnarray}
\end{subequations}
where the integer $n$ takes the values
\begin{subequations}\label{eq:n-range-Nodd-Neven}
\begin{eqnarray}
\label{eq:n-range-Nodd}
-(N-1)/2 \leq n\leq (N-1)/2\,, ~~~ \text{for odd} ~~ N\geq 1   \,,
\end{eqnarray}
and
\begin{eqnarray}
\label{eq:n-range-Neven}
-(N/2)+1 \leq n\leq N/2\,, ~~~ \text{for even} ~~ N\geq 2    \,,
\end{eqnarray}
\end{subequations}
with $N=L/a$ the number of links in the compact $4$-direction.
The momentum component in the $4$-direction is given by
\begin{equation}
p_4= 2\pi n_4/L.
\end{equation}

Using the Fourier expansion \eqref{eq:psi-psibar-Fourier-expansions}
of the fermionic field $\psi(x)$, we expand the operator $X(x)$,
defined by \eqref{eq:def-X}
in terms of $D_W$ from \eqref{eq:wilson-dirac-operator},
in the following way:
\begin{eqnarray}
X(x)\, \psi(x) &=&
X \, \frac{1}{L} \,\sum_n \psi_n(x^1,x^2,x^3)\, e^{2\pi i n x^4/L}
\nonumber \\[1mm]
&=& \frac{1}{L} \, \sum_n  e^{2\pi i n x^4/L}\, X^{(n)}(x) \,
\psi_n (x^1,x^2,x^3)\,,
\end{eqnarray}
with
\begin{equation}\label{eq:def-Xn}
X^{(n)}
\equiv
\cos(2\pi n/N) - a\, \mathbb{D}_W  -i \gamma_4\, \sin(2\pi n/N)\,,
\end{equation}
and
\begin{equation}
\mathbb{D}_W
\equiv
\frac{1}{2} \sum_{i=1}^{3}
\Big[\gamma_i(\nabla_i +\nabla_i^\ast)
+ sa\nabla_i \nabla_i^\ast\Big]\,.
\label{D_W}
\end{equation}
This operator $\mathbb{D}_W$ still contains the standard $4\times 4$ Dirac matrices $\gamma_i$.

For the gauge-field configurations (\ref{eq:condition-links}),
the operator $V$, defined by (\ref{eq:def-V}), acts on
the fermionic field in the following way:
\begin{eqnarray}
\hspace*{-8mm}
V \psi(x) &=&
V \frac{1}{L}\sum_n \psi_n(x^1,x^2,x^3)\, e^{2\pi i n x^4/L}
\nonumber \\[1mm]
\hspace*{-8mm}
&=& \frac{1}{L} \sum_n  e^{2\pi i n x^4/L} \int^\infty _{-\infty} \frac{dt}{\pi}\, X^{(n)}\,\left(t^2 + {X^{(n)}}^\dagger X^{(n)}\right)^{-1} \psi_n (x^1,x^2,x^3)
\nonumber \\[1mm]
\hspace*{-8 mm}
&\equiv &
\frac{1}{L} \sum_n  e^{2\pi i n x^4/L}\, V^{(n)}(x)\, \psi_n (x^1,x^2,x^3). \label{expand V in fourier modes}
\end{eqnarray}

We now write the fermionic action $S_F$ in terms of the
Fourier modes from \eqref{eq:psi-psibar-Fourier-expansions},
\begin{eqnarray}
\hspace*{-14mm}
&&
S_F[\overline{\psi} , \psi , U ] =
a^4 \sum_x  ~\overline{\psi}(x) D[U(x)] \psi(x) ,
\nonumber \\[1mm]
\hspace*{-14mm}
&&
=
\frac{1}{L^2}\, \sum_{m,n}   a^4 \sum_x  ~\overline{\psi}_{m}(x^1,x^2,x^3)\,
e^{-2\pi i m x^4/L}\,  D[U(x)]\, \psi_{n}(x^1,x^2,x^3)\,e^{2\pi i n x^4/L}, \nonumber \\[1mm]
\hspace*{-14mm}
&&
=
\frac{1}{L^2}\,\sum_{m,n}  a^4 \sum_x  ~\overline{\psi}_{m}(x^1,x^2,x^3)\,
 e^{2\pi i (n-m) x^4/L}
\,D^{(n)}[U(x)]\, \psi_{n}(x^1,x^2,x^3)\,,
\nonumber \\[1mm]
\hspace*{-14mm}
&&
\label{expand S_F in fourier modes1}
\end{eqnarray}
with the modes of the Ginsparg--Wilson operator $D^{(n)}$
defined by
\begin{equation}\label{eq:Ginsparg-Wilson-operator-modes-Dn}
D^{(n)} \equiv \frac{1}{a} \left( \id - V^{(n)}\right)\,,
\end{equation}
where $V^{(n)}$ follows from \eqref{eq:def-Xn}
and \eqref{expand V in fourier modes}.
In the last expression of (\ref{expand S_F in fourier modes1}),
the quantity $e^{2\pi i n x^4/L}$ is a complex number which commutes with $D^{(n)}[U(x)]$,
so that we can rewrite the above equation as follows:
\begin{eqnarray}
\hspace*{-15mm}
&&
S_F[\overline{\psi} , \psi , U ] =
\frac{1}{L^2}\,\sum_{n,m}   a^4 \sum_x
\nonumber\\[1mm]
\hspace*{-15mm}
&&
\times\,
\left( \overline{\psi}_{m}(x^1,x^2,x^3)\,e^{-2\pi i m x^4/L} \right)  D^{(n)}[U(x)] \left( \psi_{n}(x^1,x^2,x^3)\, e^{2\pi i n x^4/L} \right). \label{expand S_F in fourier modes}
\end{eqnarray}
For each value of $m$ and $n$,
we then redefine the fermionic fields as follows:
\begin{subequations}\label{eq:defs-phibarprime-phi}
\begin{eqnarray}
\overline{\psi}_{m}(x^1,x^2,x^3)\,e^{-2\pi i m x^4/L}
&\equiv& \bar{\phi^\prime}_m(x)    \,,
\\[2mm]
 \psi_{n}(x^1,x^2,x^3) \,e^{2\pi i n x^4/L}
&\equiv& \phi^\prime_n(x)  \,,
\end{eqnarray}
\end{subequations}
and rewrite the lattice fermion action as
\begin{equation}
S_F[\bar{\phi^\prime} , \phi^\prime , U ] =
\frac{1}{L^2} \sum_{n,m} a^4 \sum_x  ~
\bar{\phi^\prime}_m(x) \,D^{(n)}[U(x)]\, \phi^\prime_n(x)\,,
\end{equation}
with the operators $D^{(n)}$
from \eqref{eq:Ginsparg-Wilson-operator-modes-Dn}.

Redefining the fermionic fields again,
\begin{equation}
\psi^\prime_n (x) \equiv \frac{1}{L}\, \phi^\prime_n (x), ~~ \overline{\psi}^\prime_m (x) \equiv \frac{1}{L}\, \bar{\phi}^\prime_m (x)\,,
\end{equation}
the final action reads
\begin{eqnarray}
S_F[\overline{\psi}^\prime , \psi^\prime , U ]
&=&
\sum_{m,n} a^4 \sum_x  ~\bar{\psi^\prime}_m(x) \,D^{(n)}[U(x)] \, \psi^\prime_n(x)]
\nonumber\\[1mm]
&\equiv& \sum_{m,n}
S_F^{(m,n)}[\overline{\psi}^\prime_m , \psi^\prime_n, U ].
\end{eqnarray}
The modes $\overline{\psi}^\prime_m$  and $\psi^\prime_n$
have to satisfy the following constraints:
\begin{subequations}\label{eq:psiprime-psibarprime-constraints}
\begin{eqnarray}
\psi^\prime_n (x) &=& {\widehat{P}_{-}}^{(n)} \psi^\prime_n (x)   \,,
\\[2mm]
\overline{\psi}^\prime_m (x) &=& \overline{\psi}^\prime_m (x) P_{+}   \,,
\end{eqnarray}
\end{subequations}
with the usual projection operator $P_{+}$
and the modes of the projection operator ${\widehat{P}_{-}}$ given by
\begin{equation}
{\widehat{P}_{-}}^{(n)} =
\frac{1}{2}\left( \id - \gamma_5\, V^{(n)} \right)
\equiv
\frac{1}{2}\left( \id - \widehat{\gamma}_5^{(n)} \right).
\end{equation}
The operators $\widehat{\gamma}_5^{(n)}$ are Hermitian unitary operators. For each $n$, the operator $V^{(n)}$ is unitary and satisfies
\begin{equation}
V^{(n)\dagger}= \gamma_5 V^{(n)} \gamma_5.
\end{equation}

We now expand the Fourier modes of the fermionic fields into the following series:
\begin{subequations}\label{eq:psiprime-psibarprime-series}
\begin{eqnarray}
\psi^\prime_n (x)= \sum_j v^{(n)}_j(x)\,c_j^{(n)}   \,,
\\[2mm]
\overline{\psi}^\prime_m(x)= \sum_k \bar{c}_k^{(m)}\,\bar{v}_k^{(m)}(x)    \,.
\end{eqnarray}
\end{subequations}
Here, the $c^{(n)}$ are Grassmann-valued coefficients
and the spinor functions
$v^{(n)}_j(x)$ and $\bar{v}^{(m)}_k(x)$ form
a complete orthogonal basis of complex-valued,
$(x^1,\,x^2,\,x^3)$-antiperiodic, $(x^4)$-periodic spinors,
with the following inner products:
\begin{subequations}
\begin{equation}
\left(v^{(m)}_i, v^{(n)}_j\right)
\equiv
a^4 \sum_x v^{(m)\dagger}_i(x)\, v^{(n)}_j(x)
= \delta_{ij} \, \delta_{mn}\,,
\end{equation}
\begin{equation}
\left(\bar{v}^{(m)}_k ,\bar{v}^{(n)}_l\right)
\equiv
a^4 \sum_x \bar{v}^{(n)}_k(x)\, \bar{v}^{(m)\dagger}_l (x)
= \delta_{kl}\, \delta_{mn}\,.
\end{equation}
\end{subequations}
The spinor functions $v^{(n)}_j(x)$ and $\bar{v}^{(m)}_k(x)$
have an $x^4$-dependence given by, respectively,
$e^{2\pi i n x^4/L}$ and $e^{-2\pi i m x^4/L}$,
which traces back to the definitions \eqref{eq:defs-phibarprime-phi}.
With these expressions, the effective action for the gauge field can be factorized as follows:%
\begin{eqnarray}\label{eq:factorized-integral-lattice}
\hspace*{-8mm}
&&
\exp\left(-\Gamma[U]\right)=
\nonumber\\[1mm]
\hspace*{-8mm}
&&
\prod_{m,n} \frac{1}{Z_{m,n}^{\prime\prime}}
\Bigg[\int \prod_k d \bar{c}^{(m)}_k \prod_j dc^{(n)}_j  \exp \left(-\sum_{j,k}\bar{c}^{(m)}_k M^{(m,n)}_{kj} c_j^{(n)}\right)
\Bigg]\,,
\end{eqnarray}
in terms of the matrices
\begin{eqnarray}\label{eq:Mn-matrices}
M^{(m,n)}_{kj} [U]=
a^4 \sum_x \bar{v}^{(m)}_k(x)D^{(n)}[U(x)]v^{(n)}_j(x;U)   \,.
\end{eqnarray}
The constants $Z_{m,n}^{\prime\prime}$
in \eqref{eq:factorized-integral-lattice}
normalize the integrals, so that $\Gamma[\id]=0$.

After the Grassmann integrations
in \eqref{eq:factorized-integral-lattice},
we get the following expression for the effective action:
\begin{equation}
\Gamma[U]= - \sum_{m,n} \ln \left( \frac{1}{Z_{m,n}^{\prime\prime}} \det M^{(m,n)}_{kj} [U]  \right). \label{effective action after grassmann integration}
\end{equation}

\subsubsection{Change of the effective action under CPT}
\label{subsubsec:Change-effective-action-under-CPT}

Unlike the chiral gauge theory of the continuum,
the chiral projector \eqref{eq:projection-operator-Luscher-Phat}
for the left-handed
fermion in lattice chiral gauge theory depends on the link variables,
as follows from the definition
$\widehat{\gamma_5}[U] \equiv \gamma_5\,V[U]$.
If the gauge field is CPT transformed, the basis of the chiral fermions ${v_j}$ changes. This transformation affects the integration measure
and the effective action is CPT noninvariant. The details are as follows.

For the link configurations as considered in \eqref{eq:condition-links},
the CPT-transformed link variables are given by
\begin{equation}
U^\theta_4 = \id\,, \quad
U^\theta_i = U^\dagger_i(x-a\, \widehat{i}\,)\, ,
\end{equation}
for $i=1,\,2,\,3$ and with the unit vector $\widehat{i}$ in the $i$-direction.
Let $\mathcal{R}$ be the coordinate-reflection operator of the three coordinates $ \vec{x}\equiv (x^1,x^2,x^3)$,
\begin{equation}
\mathcal{R} : \vec{x}\rightarrow -\vec{x}\,,
\end{equation}
and let $\mathcal{R}^4$ be the coordinate-reflection operator in the fourth direction,
\begin{equation}
\mathcal{R}^4 : (\vec{x},\, x^4)\rightarrow (\vec{x},\,  -x^4).
\end{equation}
The operator $\mathbb{D}_W$, defined by (\ref{D_W}), has then the following behavior under a CPT transformation:
\begin{equation}
\mathcal{R}\mathcal{R}^4\gamma_5\,  \mathbb{D}_W [U^\theta]
\, \gamma_5\mathcal{R}^4\mathcal{R} = \mathbb{D}_W [U].
\end{equation}
The Ginsparg-Wilson-operator modes $D^{(m)}$
from \eqref{expand S_F in fourier modes1} transform as follows:
\begin{equation}
\mathcal{R}\mathcal{R}^4\gamma_5\,  D^{(n)} [U^\theta]\,
\gamma_5\mathcal{R}^4\mathcal{R} = D^{(-n)} [U].
\end{equation}

The matrices $M^{(m,n)}_{k,j}[U]$, defined by
\eqref{eq:Mn-matrices}, now change as follows
under the CPT transformation $U \rightarrow U^\theta$:
\begin{eqnarray}\label{transformation}
\hspace*{-10mm}&&
M^{(m,n)}_{k,j} [U^\theta]= a^{4} \sum_x \bar{v}^{(m)}_k(x)\, D^{(n)}[U^\theta(x)]\,v^{(n)}_j(x;U^\theta)
\nonumber\\[1mm]
\hspace*{-10mm}&&
= a^{4} \sum_x \bar{v}^{(m)}_k(x)\, \mathcal{R}\mathcal{R}^4\gamma_5 D^{(-n)}[U(x)]\gamma_5\mathcal{R}^4\mathcal{R}\, v^{(n)}_j(x;U^\theta)
\nonumber\\[1mm]
\hspace*{-10mm}&&
= \sum_{l,i} (\bar{\mathcal{Q}}_\theta^{(-m)})_{kl}\left(a^{4} \sum_x \bar{v}^{(-m)}_l(x)\, D^{(-n)}[U(x)]\, v^{(-n)}_i(x;U)\right) (\mathcal{Q}_\theta^{(-n)})_{ij}
\nonumber\\[1mm]
\hspace*{-10mm}&&
= \sum_{l,i} (\bar{\mathcal{Q}}_\theta^{(-m)})_{kl}\,
M_{li} ^{(-m,-n)}[U]\,  (\mathcal{Q}_\theta^{(-n)})_{ij}\, .
\end{eqnarray}
Here, the unitary matrices
\begin{subequations}\label{eq:def-calQ-calQbar}
\begin{eqnarray}
(\mathcal{Q}_\theta^{(-n)})_{ij} &=& a^{4} \sum_x {v^{(-n)}_j}^\dagger(\vec{x};U)\, \gamma_5\mathcal{R}^4\mathcal{R}\, v^{(n)}_j(x;U^\theta)   \,,
\\[2mm]
(\bar{\mathcal{Q}}_\theta^{(-m)})_{kl} &=& a^{4} \sum_x {\bar{v}^{(m)}_k}(x)\, \mathcal{R}\mathcal{R}^4\gamma_5 \, \bar{v}^{(-m)}_l(x) \,,
\end{eqnarray}
\end{subequations}
are obtained by introducing the projection operator $P_{+}$ and making use of the fact that
\begin{equation}
\gamma_5\, D^{(n)} = D^{(n)}\, \widehat{\gamma}_5^{(n)}\,.
\end{equation}
With the completeness of the bases $v_j^{(n)}$ and $\bar{v}_k^{(m)}$, the summation kernels of the projection operators $\widehat{P}^{(n)}_{-}$ and $P_{+}$ are
\begin{subequations}
\begin{equation}
\widehat{P}^{(n)}_{-} (x,y)
= \sum_i v_i^{(n)} (x;U)\, {v_i^{(n)}}^\dagger (y;U)
\end{equation}
and
\begin{equation}
P_{+} \frac{1}{a^4} \delta_{xy}
= \sum_l {\bar{v}_l ^{(m)\dagger}} (x)\, \bar{v}_l^{(m)} (y).
\end{equation}
\end{subequations}

The transformation (\ref{transformation}) can be absorbed by a redefinition of the fermionic variables in the multiple integral (\ref{eq:factorized-integral-lattice}), but the integration measure picks up a Jacobian factor. Under a CPT transformation,
the effective gauge-field action changes to
\begin{equation}
\Gamma[U^\theta]= \Gamma[U] - \sum_{n,m^\prime} \ln \det \left(\sum_l \left(\mathcal{Q}_\theta^{(-n)}[U]\right)_{kl} \left(\bar{\mathcal{Q}}_\theta^{(-m^\prime)}\right)_{lm}\right). \label{change in the effective action under cpt transformation}
\end{equation}
The determinants of the transformation matrices $\mathcal{Q}_\theta^{(-n)}$ depend on the link variable $U_i(x)$, which opens up the possibility
that the effective action is CPT noninvariant.

%%\newpage%%tmp
\subsection{CPT anomaly}
\label{subsec:CPT anomaly}

In this subsection, we discuss
the change of the effective gauge-field action under a CPT transformation.
But, in order to calculate the explicit expression for the CPT-violating term, we need to know the explicit form of the bases $v^{(n)}_j$
and $ \bar{v}^{(m)}_j$.

\subsubsection{Basis spinors}
\label{subsubsec:Basis-spinors}

The basis spinors for the antifermions are given by
\begin{equation}
\bar{v}^{(m)}_j (x) = \big(\bar{\xi}^{(m)}_k (x), 0\big)\,,
\end{equation}
where $\bar{\xi}^{(m)}_k (x)$ form an orthonormal basis of two-spinors in four spacetime dimensions with the explicit $x^4$-dependence
$e^{-2\pi i m x^4/L}$.

The basis vectors $v^{(n)}_j (x; U)$ are more difficult to obtain.
We have to find unitary operators $\mathcal{U}^{(n)}$ with the property
\begin{equation}\label{eq:Un-property}
\mathcal{U}^{(n)\dagger} \, \widehat{\gamma}^{(n)}_5\, \mathcal{U}^{(n)} = \gamma_5,
\end{equation}
for
\begin{equation}
\widehat{\gamma}^{(n)}_5 \equiv H^{(n)}\left( H^{(n)2} \right)^{-1/2}.
\end{equation}
Here, the Hermitian operators $H^{(n)}$ are given by
\begin{subequations}\label{eq:Hn-nmathring-noverstar-ti-wi}
\begin{eqnarray}
\label{eq:Hn}
H^{(n)}
&\equiv&
 \gamma_5
\left( \overstarn - a\, \mathbb{D}_W
 - i \gamma_4\, \mathring{n}
\right)
\nonumber\\[1mm]
&=&
\left(
\begin{matrix}
\overstarn + \frac{1}{2} \sum_{i=1}^3 w_i [U] & \mathring{n} - \frac{1}{2}\sum_{i=1}^3 \sigma_i\, t_i [U]\\[2mm]
\mathring{n} + \frac{1}{2}\sum_{i=1}^3 \sigma_i\, t_i [U] & -(\overstarn+ \frac{1}{2}\sum_{i=1}^3 w_i [U]) \\
\end{matrix}
\right)   \,,
\end{eqnarray}
with
\begin{eqnarray}
\label{eq:nmathring-noverstar}
\mathring{n} &\equiv& \sin(2\pi n/N),
\quad
\overstarn \equiv \cos(2\pi n/N)
\\[2mm]
\label{eq:ti-wi}
t_i [U] &\equiv& a\, (\nabla_i + \nabla^\ast_i),
\quad
w_i [U] \equiv a^2\, {\nabla}_i \nabla^\ast_i   \,,
\end{eqnarray}
\end{subequations}
The four-component basis spinors are then constructed as
\begin{subequations}
\begin{equation}
v^{(n)} _j (x) = \mathcal{U}^{(n)} [U]\, \chi^{(n)}_j (x)\,,
\end{equation}
with
\begin{equation}
\chi_j (x)=
\left(
\begin{matrix}
0 \\
\xi^{(n)}_j(x)  \\
\end{matrix}
\right)\,,
\end{equation}
\end{subequations}
where $\xi^{(n)}_j(x)$ form an orthonormal basis of two-spinors in four spacetime dimensions with the explicit $x^4$-dependence
$e^{2\pi i n x^4/L}$.

For the case of an odd number $N$ of links in the $x^4$ direction
(assuming odd $N\geq 3$),
we divide the domain of calculation into three subsets:
$n<0$, $n>0$, and $n=0$.
A particular property of $\widehat{\gamma}_5^{(n)}$,
\begin{equation}
\widehat{\gamma}_5^{(n)}\, \widetilde{\Gamma}_4
=-\widetilde{\Gamma}_4\, \widehat{\gamma}_5^{(-n)}, \label{tildegamma4}
\end{equation}
with the definition
\begin{equation}
\widetilde{\Gamma}_4 \equiv i\gamma_4\gamma_5\, ,
\end{equation}
suggests to impose the following condition:
\begin{equation}
 \mathcal{U}^{(-n)} [U] =
\widetilde{\Gamma}_4\,   \mathcal{U}^{(n)} [U]\,  \widetilde{\Gamma}_4\,,
\end{equation}
where the link variable $U$ on both sides of this last equation
refers to the same configuration.

\subsubsection{Fixing the phases}
\label{subsubsec:Fixing the phases}

We now obtain the required diagonalization operators
for \eqref{eq:Un-property},
first for nonzero $n$ and then for $n=0$.

In the $n\neq0$ sector, the diagonalization
operator $ \mathcal{U}^{(n)}$ is of the form
\begin{eqnarray}
\mathcal{U}^{(n)}
&=& \frac{1}{2}
\left(
\begin{matrix}
\id+ W^{(n)}&\id- W^{(n)} \\
\id- W^{(n)} & \id+ W^{(n)}  \\
\end{matrix}
\right)\frac{1}{2}
\left(
\begin{matrix}
\id+ Y^{(n)}& i(\id- Y^{(n)\dagger}) \\
i(\id- Y^{(n)}) & \id + Y^{(n)\dagger}  \\
\end{matrix}
\right)
\nonumber\\[1mm]
&&
\cdot\,
\left(
\begin{matrix}
Q_1^{(n)}&0 \\
0& Q_1^{(-n)}  \\
\end{matrix}
\right)\,,
\label{diagonalnonzero}
\end{eqnarray}
with the unitary operators
\begin{subequations}\label{eq:Wn-Yn}
\begin{eqnarray}\label{eq:Wn}
W^{(n)} &\equiv& \left(\overstarn- a\, D^{3D}_W \right)\,
\left[\left(\overstarn- a\, D^{3D}_W\right)^\dagger\left(\overstarn
- a\, D^{3D}_W\right)\right]^{-1/2},
\\[2mm]
\label{eq:Yn}
Y^{(n)} &\equiv&
\left[\left(\overstarn- a\, D^{3D}_W\right)^\dagger W^{(n)} +i \mathring{n}\right]\,
\left[\left(\overstarn- a\, D^{3D}_W\right)^\dagger \left(\overstarn- a\, D^{3D}_W\right) + \mathring{n}^2\right]^{-1/2},
\nonumber\\[1mm]
&&
\end{eqnarray}
\end{subequations}
and
\begin{eqnarray}
\label{eq:DW3D}
D^{3D}_W &\equiv&  \frac{1}{2} \sum_{i=1}^{3} \Big(\sigma_i(\nabla_i+\nabla_i^\ast)+ sa\nabla_i^\ast \nabla_i \Big)\,.
\end{eqnarray}
One possible choice for $Q_1^{(n)}$ is
\begin{equation}
Q_1^{(n)} [U] =
\left\{\begin{array}{cc}
\id\,,         & \;\;\text{for}\;\;n>0\,, \\[1mm]
W^{(n)}[U]^\dagger\,, &  \;\;\text{for}\;\;n<0\,.
       \end{array}
\right.
\end{equation}
A change of $n$ to $-n$  gives
\begin{equation}
W^{(-n)}= W^{(n)}, ~~ Y^{(-n)}= Y^{(n)\dagger} \,.
\label{changeinn}
\end{equation}

In the $n=0$ sector, the diagonalization operator $ \mathcal{U}^{(n)}$
is of the form
\begin{equation}
\mathcal{U}^{(0)}= \frac{1}{2}
\left(
\begin{matrix}
\id+ W^{(0)^\dagger}&\id- W^{(0)} \\
-\id+ W^{(0)^\dagger} & \id+ W^{(0)}  \\
\end{matrix}
\right)\,, \label{diagonal0}
\end{equation}
with $ W^{(0)}$ defined by \eqref{eq:Wn} for $n=0$.
As discussed in App.~B of Ref.~\cite{KlinkhamerSchimmel2002},
other possible choices for $\mathcal{U}^{(0)}$ are
characterized by an integer $k^{(0)}\in\mathbb{Z}$ and give an additional
factor $(2\,k^{(0)}+1)$ in the final
result \eqref{eq:Gamma-CS-like-nonperturbative}.

\subsubsection{CPT anomaly for odd $N\geq 3$}
\label{subsubsec:CPT-anomaly-for-odd-N}

The diagonalization operators $\mathcal{U}^{(n)} [U]$ are given by (\ref{diagonalnonzero}) and (\ref{diagonal0}) and the CPT-violating factor can be calculated as follows.

The operator $D^{3D}_W$ from \eqref{eq:DW3D} transforms under CPT as
\begin{equation}
D^{3D}_W [U^\theta] = \mathcal{R}\mathcal{R}^4 ~D^{3D}_W [U]^\dagger~ \mathcal{R}^4\mathcal{R}.
\end{equation}
The operators $W^{(n)}$ and $Y^{(n)}$ transform under CPT  as follows:
\begin{subequations}\label{eq:Wn-Yn-transf}
\begin{eqnarray}
\label{eq:Wn-transf}
W^{(n)} [U^\theta] &=& \mathcal{R}\mathcal{R}^4 ~W^{(n)^\dagger}[U]~ \mathcal{R}^4\mathcal{R}   \,,
\\[2mm]
\label{eq:Yn-transf}
Y^{(n)} [U^\theta] &=& \mathcal{R}\mathcal{R}^4 ~W^{(n)} [U]
 \,Y^{(n)} [U]\, W^{(n)^\dagger}[U]~ \mathcal{R}^4\mathcal{R}   \,.
\end{eqnarray}
\end{subequations}
With the help of \eqref{eq:Wn-transf} and \eqref{eq:Yn-transf},
we calculate the
changes of the diagonalization operators $\mathcal{U}^{(n)}$
under a CPT transformation for $n<0$, $n>0$, and $n=0$.
The results are for $n<0$:   %%frk: better style??
\begin{subequations}\label{eq:Un-CPTtransformed-all-n}
\begin{equation}
\mathcal{R}\mathcal{R}^4\gamma_5 ~\mathcal{U}^{(n)} [U^\theta]~ \gamma_5 \mathcal{R}^4\mathcal{R} =
\widetilde{\Gamma}_4\, \mathcal{U}^{(n)} [U]\, \widetilde{\Gamma}_4 ~
\left(
\begin{matrix}
Y^{(n)} &0 \\
0& W^{(n)} Y^{(n)^\dagger} W^{(n)^\dagger} \\
\end{matrix}
\right),
\end{equation}
for $n>0$:
\begin{equation}
\mathcal{R}\mathcal{R}^4\gamma_5 ~\mathcal{U}^{(n)} [U^\theta]~ \gamma_5 \mathcal{R}^4\mathcal{R} =
\widetilde{\Gamma}_4\, \mathcal{U}^{(n)} [U]\, \widetilde{\Gamma}_4 ~
\left(
\begin{matrix}
 W^{(n)} Y^{(n)} W^{(n)^\dagger}&0 \\
0& Y^{(n)^\dagger} \\
\end{matrix}
\right),
\end{equation}
and for $n=0$:
\begin{equation}
\hspace*{-0mm}
\mathcal{R}\mathcal{R}^4\gamma_5 ~\mathcal{U}^{(0)} [U^\theta]~ \gamma_5 \mathcal{R}^4\mathcal{R} = \widetilde{\Gamma}_4 \mathcal{U}^{(0)} [U] \widetilde{\Gamma}_4.
\end{equation}
\end{subequations}
The changed transformation matrices are for $n=0$:
\begin{subequations}
\begin{eqnarray}
\hspace*{-7mm}
\left(\mathcal{Q}^{(0)}_\theta[U]\right)_{ij} &=& a^4 \sum_x \chi^{(0)\dagger}_i(x) \mathcal{U}^{(0)} [U]^\dagger  \mathcal{R}\mathcal{R}^4\gamma_5 ~\mathcal{U}^{(0)} [U^\theta]~ \chi^{(0)}_j(x) \nonumber\\[1mm]
\hspace*{-7mm}
&=& a^4 \sum_x \chi^{(0)\dagger}_i(x) \mathcal{U}^{(0)} [U]^\dagger \mathcal{U}^{(0)} [U^\theta]~\mathcal{R}^4\mathcal{R}\gamma_5 ~\chi^{(0)}_j(x)\,,
\end{eqnarray}
for $n>0$:
\begin{eqnarray}
\hspace*{-13mm}
&&
\left(\mathcal{Q}^{(n)}_\theta[U]\right)_{ij}
\nonumber\\[1mm]
\hspace*{-13mm}
&&
= a^4 \sum_x \left(0, \xi^{(n)\dagger}_i(x)\right)  \left(
\begin{matrix}
 W^{(n)} Y^{(n)} W^{(n)^\dagger}&0 \\
0& Y^{(n)^\dagger} \\
\end{matrix}
\right) \mathcal{R}\mathcal{R}^4\gamma_5 ~ \left(
\begin{matrix}
0 \\
\xi^{(n)}_j(x)  \\
\end{matrix}
\right)\,,
\end{eqnarray}
and for $n<0$:
\begin{eqnarray}
\hspace*{-15mm}
&&
\left(\mathcal{Q}^{(n)}_\theta[U]\right)_{ij}
\nonumber\\[1mm]
\hspace*{-15mm}
&&
= a^4 \sum_x \left(0, \xi^{(n)\dagger}_i(x)\right)  \left(
\begin{matrix}
Y^{(n)} &0 \\
0& W^{(n)} Y^{(n)^\dagger} W^{(n)^\dagger} \\
\end{matrix}
\right) \mathcal{R}\mathcal{R}^4\gamma_5 ~ \left(
\begin{matrix}
0 \\
\xi^{(n)}_j(x)  \\
\end{matrix}
\right).
\end{eqnarray}
\end{subequations}
We shall later see that the transformation matrices for the $n<0$ modes
and the $n>0$ modes do not contribute to the final expression of the anomalous term.

The changed transformation matrices $\bar{\mathcal{Q}}^{(m^{\prime})}_\theta[U]$
are the same for all values of the Fourier index $m^{\prime}$:
\begin{equation}
\left(\bar{\mathcal{Q}}^{(m^{\prime})}_\theta[U]\right)_{kl} = \left( \bar{\xi}^{(m^{\prime})}_k(x), 0\right)~ \mathcal{R}\mathcal{R}^4\gamma_5 \left(
\begin{matrix}
\bar{\xi}^{(m^{\prime})^\dagger}_l(x)  \\
0 \\
\end{matrix}
\right).
\end{equation}
The required combinations of transformation matrices give for $n=0$:
\begin{subequations}\label{eq:det-n-all}
\begin{equation}
\left(\mathcal{Q}^{(0)}_\theta[U]\right)_{kl} \left(\bar{\mathcal{Q}}^{({m^\prime})}_\theta[U]\right)_{lm} = -a^4 \sum_x \xi^{(0)\dagger}_k(x)   W^{(0)}[U]^\dagger \xi^{(0)}_m (x)
\,\delta_{m^\prime 0}\, ,\label{eq:det-n-equal-0}
\end{equation}
for $n>0$:
\begin{eqnarray}
\hspace*{-10mm}
&&
\sum_{l}\left(\mathcal{Q}^{(n)}_\theta[U]\right)_{kl} \left(\bar{\mathcal{Q}}^{({m^\prime})}_\theta[U]\right)_{lm}
\nonumber\\[1mm]
\hspace*{-10mm}
&&
= -a^4 \sum_x \xi^{(n)\dagger}_k(x)   \left( W^{(n)}[U]Y^{(n)}[U]W^{(n)}[U]^\dagger \right) \xi^{(n)}_m (x) \,\delta_{m^\prime n}\,, \label{eq:det-n-greater-0}
\end{eqnarray}
and for $n<0$:
\begin{equation}
\sum_{l}\left(\mathcal{Q}^{(n)}_\theta[U]\right)_{kl} \left(\bar{\mathcal{Q}}^{({m^\prime})}_\theta[U]\right)_{lm} = -a^4  \sum_x \xi^{(n)\dagger}_k(x)   Y^{(n)}[U]^\dagger \xi^{(n)}_m (x)
\,\delta_{m^\prime n}\,. \label{eq:det-n-less-0}
\end{equation}
\end{subequations}
For the derivation of (\ref{eq:det-n-all}), we have used
\begin{subequations}
\begin{equation}
\bar{\xi}^{(m^\prime)}_k = \xi^{(m^\prime)\dagger}_k(x)
\end{equation}
and the completeness relation of the two-spinor basis $\xi^{(n)}_k (x)$,
\begin{equation}
\sum_{k} \xi^{(m^\prime)\dagger}_k(x)  \xi^{(n)}_k (y) = a^{-4} ~\id ~\delta_{xy}~  \delta_{m^\prime n}.
\end{equation}
\end{subequations}

Because $W^{(n)}$ and $Y^{(n)}$ are unitary, the determinant of (\ref{eq:det-n-greater-0}) for $n>0$ is the inverse of the
determinant  of (\ref{eq:det-n-less-0}) for $n<0$, where we have used the
relations (\ref{changeinn}). This gives
\begin{eqnarray}
&&
\prod_{n>0} \prod_{m^\prime}
\det \left( \sum_l \left(\mathcal{Q}^{(n)}_\theta[U]\right)_{kl} \left(\bar{\mathcal{Q}}^{(m^\prime)}_\theta[U]\right)_{lm}\right)
\nonumber\\[1mm]
&&
\times\,
\det \left(\sum_l \left(\mathcal{Q}^{(-n)}_\theta[U]\right)_{kl} \left(\bar{\mathcal{Q}}^{(m^\prime)}_\theta[U]\right)_{lm}\right) = 1. \label{cancellation of terms for odd N}
\end{eqnarray}
We see from (\ref{cancellation of terms for odd N})
that the anomalous terms arising from positive frequencies ($n>0$) are canceled by the terms arising from negative frequencies ($n<0$), so that only the $n=0$ term survives. This $n=0$ term
is given by \eqref{eq:det-n-equal-0}, which effectively sets
$m'=0$.

To summarize, the change in the effective gauge-field action
under a CPT transformation is, for odd $N \geq 3$, given by
\begin{equation} \label{eq:change-under-CPT-odd-N}
\Delta \Gamma [U] \equiv  \Gamma [U^\theta]- \Gamma [U] = - \ln \det \left(a^4 \sum_x \xi^{(0)\dagger}_k(x)   W^{(0)}[U]^\dagger \xi^{(0)}_m (x)\right),
\end{equation}
with the unitary operator
\begin{equation}\label{eq:def-W0-U}
W^{(0)}[U] = \left(\id - a D_W^{3D} [U]\right)
\left[\left(\id - a D_W^{3D} [U]\right)^\dagger \left(\id
- a D_W^{3D} [U]\right)\right]^{-1/2}.
\end{equation}

\subsubsection{CPT anomaly for even $N\geq 4$}
\label{subsubsec:CPT-anomaly-for-even-N}

For even $N$ (equal to or larger than $4$),
we divide the Fourier modes $n$ into four subsets:
$-N/2 < n<0$, $n=0$, $0<n<N/2$, and $n=N/2$.
The case $N=2$, for $x^4$-independent gauge fields,
has already been discussed in Ref.~\cite{KlinkhamerSchimmel2002}.

Equation (\ref{tildegamma4}) is also valid for even $N$,
as long as $n\neq N/2$. For $n=N/2$, we have
\begin{equation}
\widehat{\gamma}_5^{(N/2)}\, \widetilde{\Gamma}_4
=-\widetilde{\Gamma}_4\, \widehat{\gamma}_5^{(N/2)}.
\end{equation}
Hence, the results from Sec.~\ref{subsubsec:CPT-anomaly-for-odd-N} can be used for $n\neq N/2$.
But the $n=N/2$ diagonalization operator needs to be investigated separately. 
For $n=N/2$, we have
\begin{equation}
\mathcal{U}^{(N/2)}= \frac{1}{2}
\left(
\begin{matrix}
\id+ W^{(N/2)^\dagger}&\id- W^{(N/2)} \\
-\id+ W^{(N/2)^\dagger} & \id+ W^{(N/2)}  \\
\end{matrix}
\right)\,, \label{diagonalN/2}
\end{equation}
where  the unitary operator $W^{(N/2)}[U]$ is defined as
\begin{equation}\label{eq:def-WNoverTwo-U}
W^{(N/2)}[U] \equiv -\left(\id + a D_W^{3D} [U]\right)
\left[\left(\id + a D_W^{3D} [U]\right)^\dagger
\left(\id + a D_W^{3D} [U]\right)\right]^{-1/2}.
\end{equation}

The total change in effective gauge-field action under a CPT transformation
is, for even $N \geq 4$, determined by
\begin{eqnarray}\label{eq:total-change-under-CPT-even-N}
&&
\det \left( \sum_l \left(\mathcal{Q}^{(0)}_\theta[U]\right)_{kl} \left(\bar{\mathcal{Q}}^{(0)}_\theta[U]\right)_{lm}\right)
\nonumber\\[1mm]
&&
\times\,
\det \left(\sum_l \left(\mathcal{Q}^{(N/2)}_\theta[U]\right)_{kl} \left(\bar{\mathcal{Q}}^{(N/2)}_\theta[U]\right)_{lm}\right)
\nonumber\\[1mm]
&&
=
\det \left(a^4 \sum_x \xi^{(0)\dagger}_k(x)   W^{(0)}[U]^\dagger \xi^{(0)}_m (x)\right)
\nonumber\\[1mm]
&&
\times\,
\det \left(a^4 \sum_x \xi^{(N/2)\dagger}_k(x)   W^{(N/2)}[U]^\dagger \xi^{(N/2)}_m (x)\right)\,,
\end{eqnarray}
with the unitary operators $W^{(0)}$ and $W^{(N/2)}$ given by, respectively,
\eqref{eq:def-W0-U} and \eqref{eq:def-WNoverTwo-U}.

The expressions \eqref{eq:change-under-CPT-odd-N} for odd $N \geq 3$
and \eqref{eq:total-change-under-CPT-even-N} for even $N \geq 4$ give
the change of the effective gauge-field action under
a CPT transformation according
to \eqref{change in the effective action under cpt transformation}
and are the main results of the nonperturbative lattice calculation.
In order to better understand the meaning of these expressions,
we consider the continuum limit of them in the next subsection.

%%\newpage%%tmp
\subsection{CPT anomaly in the continuum limit}
\label{subsec:Continuum-limit}

As mentioned in Sec.~\ref{subsec:Lattice-setup},
we first consider an Abelian $U(1)$ gauge field
coupled to a single unit-charge chiral fermion.
The change in the effective gauge-field
action under a CPT transformation for an odd number $N$ of links in the 4-direction depends only on $W^{(0)}[U]$, see (\ref{eq:change-under-CPT-odd-N}).
For an even number $N$ of links in the 4-direction, the corresponding change is given by (\ref{eq:total-change-under-CPT-even-N}).

Consider an even number $N$ of links in the 4-direction and
introduce the following short-hand notations:
\begin{equation}
W^{(-)^\dagger}  \equiv W^{(0)^\dagger} , ~~~  W^{(+)^\dagger}  \equiv W^{(N/2)\dagger} ,
\end{equation}
with
\begin{eqnarray}
W^{(\pm)^\dagger} &=& \mp \big(\id \pm a D^{3D}_W\big)^\dagger \,
\big[
\big(\id \pm a D^{3D}_W\big)\,\big(\id \pm a D^{3D}_W\big)^\dagger \big]^{-1/2} \nonumber \\[2mm]
&=& - \big( D^{3D}_W \pm 1/a\big)^\dagger \,
\big[
\big(D^{3D}_W \pm 1/a\big)\,\big(D^{3D}_W \pm 1/a\big)^\dagger
\big]^{-1/2}
\end{eqnarray}
for $D^{3D}_W$ from \eqref{eq:DW3D}.
The change in the effective gauge-field action is calculated
from \eqref{eq:total-change-under-CPT-even-N}  as
\begin{subequations}\label{eq:continuouslimit}
\begin{eqnarray}
\label{eq:continuouslimit-full}
\hspace*{-10mm}
\bigtriangleup \Gamma [U] &=& i \left( \text{Im} \lbrace\ln \det ( D^{3D}_W - 1/a ) \rbrace + \text{Im} \lbrace\ln \det ( D^{3D}_W + 1/a ) \rbrace \right)
 \\[2mm]
\label{eq:continuouslimit-short-hand}
\hspace*{-10mm}
&\equiv& i \left( \text{Im} \lbrace\ln \det (D - m_{+} ) \rbrace + \text{Im} \lbrace\ln \det ( D - m_{-}) \rbrace \right)\,,
\end{eqnarray}
\end{subequations}
where, in \eqref{eq:continuouslimit-short-hand}, we have
introduced further short-hand notations,
\begin{equation}\label{eq:defs-D-mplus-mnin}
D \equiv D^{3D}_W\,,
\quad
m_{+} \equiv 1/a\,,
\quad
m_{-} \equiv -(1/a) \,.
\end{equation}
The first operator in (\ref{eq:continuouslimit-full})
is a Wilson--Dirac operator with positive mass $1/a$ and the second operator
is a Wilson--Dirac operator with negative mass $-1/a$. Because of the antiperiodic boundary conditions in the $x^1,\,x^2,\,x^3$ directions,
the masses for these operators are effectively increased
by a contribution of order $a/(L^\prime)^2$.
The values of the positive and negative effective masses are now
\begin{subequations}\label{eq:effectivemass}
\begin{eqnarray}
\label{eq:effectivemass-plus}
 m_{+}^\text{(eff)} &=& +1/a +c_{+}\,a/(L^\prime)^2  \,,
\\[2mm]
\label{eq:effectivemass-minus}
m_{-}^\text{(eff)}  &=& -1/a +c_{-}\,a/(L^\prime)^2  \,,
\end{eqnarray}
\end{subequations}
with positive constants $c_\pm$.

The vacuum-polarization kernel of the effective gauge-field
action in three dimensions has been calculated in
Ref.~\cite{CosteLuscher1989} to second order in the bare coupling constant $e$. We adopt a similar approach, in order to calculate the change in
the effective action under a CPT transformation.

For this purpose, we consider an auxiliary theory
of a nonchiral four-component Dirac fermion field $\Psi(x)$
with the following action over the four-dimensional
lattice \eqref{eq:4D-hypercubic-lattice}:
\begin{equation}\label{eq:auxiliary-fermionic-action}
S_F = -a^4 \,\sum_x \overline{\Psi}(x)\, [D - m]\, \Psi(x)\,,
\end{equation}
where $D$ is the operator from \eqref{eq:defs-D-mplus-mnin} and
$m$ an arbitrary mass.
The corresponding effective gauge-field action $\Gamma[A]$ is given by
\begin{equation}
\Gamma[A]= \ln \det [D-m]\,.
\end{equation}
The fermion propagator $S(x,y)_{\alpha\beta}$ from
\eqref{eq:auxiliary-fermionic-action} is defined by
\begin{equation}
[(-D + m)S(x,y)]_{\alpha\beta} =
\frac{1}{a^4} \,\delta_{\alpha\beta}\,\delta_{xy}\,.
\end{equation}
In momentum space, we have
\begin{eqnarray}
S(x,y) &=&
\frac{1}{L}\,\sum_{n} \int_{-\pi/a}^{\pi/a} \frac{d^{3} p}{2\pi^3}\, e^{ip(\vec{x}-\vec{y})}\,e^{2\pi i n (x^4-y^4)/L}\, S_n (\vec{p})
\nonumber\\[1mm]
&=&
\frac{1}{L}\,\sum_{n} \int_{-\pi/a}^{\pi/a} \frac{d^{3} p}{2\pi^3}\,
e^{i\vec{p}\cdot(\vec{x}-\vec{y})}\,e^{2\pi i n (x^4-y^4)/L}\, S(p_n) \,,
\label{fourier transform of the propagator}
\end{eqnarray}
with, as before,
\begin{equation}
p_n \equiv (\vec{p},\, \rho_n )\,,\quad \rho_n \equiv 2\pi n/L.
\end{equation}
A comment on the Fourier transforms
of \eqref{fourier transform of the propagator} is in order.
The momentum steps in the fourth direction
and those in the other three directions
are, respectively, of order $1/L$ and $1/L^\prime$,
with $L^\prime \gg L$.
Hence, we have kept in \eqref{fourier transform of the propagator}
the summation for the momentum in the fourth direction
but used an integral for the momenta in the three other directions.

Next,  define a quantity $Q(p_n)$ in such a way that
\begin{equation}
S(p_n) = Q(p_n)^{-1}.
\end{equation}
This quantity $Q(p_n)$
is a function of $\widehat{p}_{n_\mu}$ and $\widetilde{p}_{n_\mu}$,
which are defined as follows:
\begin{equation}
\widehat{p}_{n_\mu} \equiv
\frac{2}{a}\, \sin \left(\frac{1}{2}\,a\,{p_n}_\mu\right), ~~
\widetilde{p}_{n_\mu}  \equiv \frac{1}{a}\, \sin (a\,{p_n}_\mu).
\end{equation}
We expand the Dirac operator $D$ in powers of the coupling constant $e$,
\begin{equation}
D= \sum_k^\infty e^k D_k \,,
\end{equation}
where,  for $k\geq 1$, we have
\begin{eqnarray}
\hspace*{-16mm}
&&D_k \,\Psi(x) =  \frac{(ia)^k}{2ak!}\sum_{i=1}^3
\nonumber\\[1mm]
\hspace*{-16mm}
&&
\times\,
[ A_i(x)^k (s+\gamma_i)\Psi(x+a\,\widehat{i}\,)+ (-1)^k A_i (x-a\,\widehat{i}\,)^k(s-\gamma_i)\Psi(x-a\,\widehat{i}\,)].
\end{eqnarray}

For the effective gauge-field action, there is the following expansion
in powers of the fermion charge:
\begin{equation}
\Gamma[A] = \sum_k^\infty e^k\, \Gamma_k[A].
\label{effective-action-as-series-in-e}
\end{equation}
With the Fourier transform of the gauge field $A_\mu$, we write
the two-point function as
\begin{equation}\label{eq:Gamma2-A}
\Gamma_2 [A] = -i\,\frac{1}{2}\,\frac{1}{L}\, \sum_{n}
\int_{-\pi/a}^{\pi/a} \frac{d^{3} q}{2\pi^3}\,
{A_i}(-q_n) \,\widehat{\pi}_{ij}(q_n)\,  {A_j}(q_n)\,,
\end{equation}
where we have included the same prefactor $-i/2$ as
in \eqref{factor1} and where
the vacuum polarization tensor
$\widehat{\pi}_{ij}(q_n)$ is now given by
\begin{eqnarray}
\widehat{\pi}_{ij}(q_n)&=&
\frac{1}{2}\,\frac{1}{L}\,
\sum_{m} \int_{-\pi/a}^{\pi/a} \frac{d^{3} p}{2\pi^3}\;
\big[1-T_0 (q_n)\big]
\nonumber\\[1mm]&&
\times \,
\text{tr} \Big\{ [Q\left(p_m+q_n/2\right)]^{-1} \partial_i Q(p_m) ~  [Q\left(p_m-q_n/2\right)]^{-1}\partial_jQ(p_m) \Big\}.
\nonumber\\[1mm]&&
\label{vacuum polarisation tensor in lattice}
\end{eqnarray}
The symbol $[1-T_0(q_n)]$ in the above equation stands for a Taylor subtraction at zero momentum.
Just as for the perturbative calculation of
Sec.~\ref{subsec:Calculation-pert}, the anomalous term originates
from the $m=0$ sector of \eqref{vacuum polarisation tensor in lattice}.
We now focus on this $m=0$ sector [denoted by the superscript `$(0)$']
and will mention later the contribution of the $m\ne 0$ terms.

In the continuum limit, we can use the three-dimensional result
from Ref.~\cite{CosteLuscher1989},
\begin{subequations}\label{eq:vacuum-polarisation-tensor-continuum-limit-defs}
\begin{eqnarray}
\label{eq:vacuum-polarisation-tensor-continuum-limit}
\hspace*{-5mm}
\widehat{\pi}_{ij}^{(0)\,\text{(cont.)}}(q_n)
&=&
\lim_{a\to 0}\, \widehat{\pi}_{ij}^{(0)}(q_n)
\nonumber\\[1mm]
\hspace*{-5mm}
&=&
\frac{1}{L}\,A(q_n^2)\, \epsilon_{ijk}\, {q_n}^k
+\frac{1}{L}\,B(q_n^2)\, (q_n^2\, \delta_{ij} - {q_n}_i\, {q_n}_j) \,,
\end{eqnarray}
with amplitudes $A({q_n}^2)$  and $B({q_n}^2)$ given by
\begin{eqnarray}
\hspace*{-5mm}
 A({q_n}^2) &=& \frac{1}{2}\,a_0 + \frac{1}{8\pi} \int_0^1 dt
~\Big\{ 1-m\,[m^2+t(1-t)\,{q_n}^2]^{-1/2} \Big\}  \,,
\\[2mm]
\hspace*{-5mm}
B({q_n}^2) &=& \frac{1}{4\pi} \int_0^1 dt
~\Big\{ 1-m\,[m^2+t(1-t)\,{q_n}^2]^{-1/2}\Big\}   \,,
\end{eqnarray}
\end{subequations}
where `$m$' is the mass defined by \eqref{eq:auxiliary-fermionic-action}
and not a Fourier component (for the moment, we have Fourier component $m=0$). Henceforth, we drop the superscript `(cont.)' of
\eqref{eq:vacuum-polarisation-tensor-continuum-limit} and
focus on the part with an odd number of momenta,
containing the Levi--Civita symbol and the
$A({q_n}^2)$ amplitude. With the Wilson parameter $s=-1$,
we have the constant $a_0 = -1/(2\pi)$.
In the large negative $m$ limit for a fixed value of ${q_n}^2$,
the odd-momentum part of the polarization tensor
$\widehat{\pi}_{ij}^{(0)}(q)$ vanishes, whereas,
in the large positive $m$ limit for fixed ${q_n}^2$,
the odd-momentum part of the polarization tensor becomes
\begin{equation}\label{eq:limit-result-pi-ij}
 \lim_{m\to\infty} \widehat{\pi}_{ij}^{(0)\text{\,(odd-mom)}}(q)
= \frac{1}{L}\,\frac{a_0}{2}~ \epsilon_{ijk} ~ q^k
= -\frac{1}{4\pi}\frac{1}{L}\, ~ \epsilon_{ijk} ~ q^k\,.
\end{equation}

As mentioned above, the anomalous contribution
\eqref{eq:limit-result-pi-ij} originates
from the $m=0$ Fourier sector
of \eqref{vacuum polarisation tensor in lattice}.
The $m\ne 0$ Fourier terms of \eqref{vacuum polarisation tensor in lattice}
contribute a further term $\propto (1/a)\,\epsilon_{ijk} ~ q^k$,
which is $L$-independent
and divergent in the continuum limit $a\to\infty$.
Just as discussed in Sec.~\ref{subsec:Calculation-pert},
this extra term can be removed by a suitable renormalization
procedure.

With the
results \eqref{eq:vacuum-polarisation-tensor-continuum-limit-defs}
and \eqref{eq:limit-result-pi-ij} obtained from the auxiliary
theory \eqref{eq:auxiliary-fermionic-action},
we now return to the original chiral gauge theory.
The first term in (\ref{eq:continuouslimit}) has a positive
mass $m=1/a$ and the second term has a negative mass  $m=-1/a$,
so that the second term does not contribute to the anomalous
change in the effective gauge-field action.
The anomalous change in the effective action
follows solely from the first term
in (\ref{eq:continuouslimit}) and is determined
by \eqref{eq:limit-result-pi-ij}.
Up till now, we have considered an even number $N$ of links in the 4-direction.
For an odd number $N$ of links, the second term in (\ref{eq:continuouslimit}) does not appear and the result is the same as for
even $N$.

Changing from momentum space to position space,
the first term in (\ref{eq:continuouslimit}) gives,
using \eqref{eq:limit-result-pi-ij},
the following result up to order $e^2$ in the effective
gauge-field action \eqref{eq:Gamma2-A}:
\begin{equation}
e^2\,\Gamma_2^\text{\,(odd-mom.)}[A] =
2\pi i\,e^2\, \frac{1}{L}
\sum_{n_4} a  \int_{\mathbb{R}^3} d^{3}x
~
\omega_\text{CS} [A(\vec{x}, n_4\,a)], \label{discretecase}
\end{equation}
where the Chern--Simons density $\omega_\text
{CS}$ has been
defined in \eqref{eq:omega-CS}.
The continuum limit has $a\to 0$ and $N \to \infty$, with
constant product  $Na=L$.

Next, change from a Euclidean metric signature to a
Lorentzian metric signature and include all fermions
of the chiral gauge theory \eqref{eq:U1theory-G-RL},
with
all of these fermions treated equally on the lattice.
The expression (\ref{discretecase}) then becomes
\begin{equation}
e^2\,\Gamma^\text{\,(odd-mom.)}_2[A] =
-F\,e^2\, \frac{2\pi}{L}
\int_0^L dx^4~\int_{\mathbb{R}^3} d^{3}x~ \omega_\text{CS} [A(\vec{x}, x^4)]\,,
\label{eq:Gamma-CS-like-nonperturbative}
\end{equation}
with an extra factor $i$ for the Lorentzian metric signature and
an overall numerical factor $F$ from \eqref{eq:sum-charges-square-F}
due to the contribution of  all chiral fermions of the theory \eqref{eq:U1theory-G-RL}.

%%\newpage%%tmp
\section{Discussion}
\label{sec:Discussion}

In this section, we present six general remarks in order to clarify the
calculations performed in Secs.~\ref{sec:Perturbative-approach}
and \ref{sec:Nonperturbative-approach}.

First, we must explain how an apparently CPT-invariant theory
has produced CPT violation.
With an extended version of the generalized Pauli--Villars regularization
for the perturbative
calculation, the regulator masses $M_r$
in \eqref{eq:full-PV-type-lagrangian}
are the source of the Lorentz and CPT violation
(these Lorentz-violating terms in the regularized action appear to be
necessary in order to maintain the gauge invariance of the second-quantized
theory, as discussed in Sec.~6 of Ref.~\cite{Klinkhamer1999}).
With the lattice regularization for the nonperturbative calculation,
the crucial observation is that the gauge-covariant diagonalization operators (\ref{diagonalnonzero}) and (\ref{diagonal0}) are not CPT invariant, as shown by (\ref{eq:Un-CPTtransformed-all-n}).

Let us expand on the CPT noninvariance of the lattice calculation.
For an odd number $N$ of links in the 4-direction,
we have explicitly shown that
the changes of the nonperturbative effective gauge-field action
under a CPT transformation for positive $n$
are canceled by the corresponding changes for negative $n$.
But the $n=0$ contribution has no counterpart to cancel its change under
a CPT transformation.
Specifically, the change of the $n=0$ diagonalization operator is given by
\begin{equation}
\mathcal{R}\mathcal{R}^4\gamma_5
~\mathcal{U}^{(0)} [U^\theta]~
\gamma_5 \mathcal{R}^4\mathcal{R} = \mathcal{U}^{(0)} [U]
\left(
\begin{matrix}
 W^{(0)^\dagger}&0 \\
0 &  W^{(0)}  \\
\end{matrix}
\right)\,,
\end{equation}
where $W^{(0)^\dagger}$ acts on left-handed fermions and $W^{(0)}$ acts on right-handed fermions. The CPT transformation leads to another theory with different basis spinors~\cite{KlinkhamerSchimmel2002}. This different theory can be transformed back to the original one by a redefinition of the spinors. But, then, the integration measure picks up a Jacobian factor and the effective gauge-field action $\Gamma[U]$ changes,
\begin{equation}
\Delta \Gamma[U] \equiv \Gamma[U^\theta] - \Gamma[U] = - \ln \det \left( a^4 \sum_x \xi^{(0)\dagger}_k (x) \left(W^{(0)} [U]\right) \xi^{(0)}_m (x)\right).
\end{equation}

For an even number $N$ of links in the 4-direction, we can give the same
argument as for an odd number of links.
The changes in the measure for $0<n<N/2$ are again canceled by the
corresponding changes for negative $n$.
The remaining factors are those for $n=0$ and $n=N/2$.
But the additional factor for $n=N/2$ is a lattice artefact and vanishes in the continuum limit.

Note also that the CPT anomaly vanishes for Dirac fermions
with both left- and right-handed components,
\begin{equation}
\ln \det W^{(0)^\dagger} + \ln \det W^{(0)} = 0.
\end{equation}

Second, let us discuss the conditions on the background gauge field.
If the gauge fields depend upon the compactified coordinate $x^4$, they should not oscillate too fast with respect to the $x^4$ coordinate.

In the perturbative approach, we Fourier expand the gauge
field $A_\mu$  in the following way:
\begin{equation}
A_\mu (x) = \frac{1}{L} \sum_{n={-\infty}}^{\infty} \int \frac{d^{3}p}{(2\pi)^3} ~ e^{2\pi i n x^4/L}~ e^{i \vec{p}\cdot \vec{x}}~ {A_\mu} (p_n)\,.
\end{equation}
The frequency of oscillation of $A_\mu$ with respect to $x^4$ is $n/L$.
The discrete momentum corresponding to the coordinate $x^4$ is given by
\begin{equation}
\rho_n =  2\pi n/L . \label{regmass}
\end{equation}
For the generalized Pauli--Villars regularization used,
the regulator mass scale $M$ must be very much larger than the
momentum component $\rho_n =  2\pi n/L$, as discussed
on the lines above \eqref{eq:trace-gammatilde-ijk}.
Hence, the condition on the gauge fields is given by
\begin{equation}\label{eq:perturbative-condition}
n \ll M \, L \,,
\end{equation}
where $n$ controls the dimensionless
oscillation frequency of the gauge field
$A_\mu$ with respect to $x^4$ and $L$ is the range of the compactified coordinate $x^4$.

In the nonperturbative approach, the 4-direction momentum $\rho_n$
of the external gauge fields must be very small
compared to the regulator scale $1/a$,
in order to be able to apply the continuum expressions
of Sec.~\ref{subsec:Continuum-limit}. There is, then, the following
condition (using $\rho_n \sim n/L$):
\begin{equation}\label{eq:nonperturbative-condition}
\frac{n}{L} = \frac{n}{Na} \ll \frac{1}{a}< m_{+} ~,
\end{equation}
where $m_{+}$ is the effective mass  \eqref{eq:effectivemass-plus}
for the Wilson--Dirac operator (this effective mass $m_{+}$
is similar to the Pauli--Villars regulator mass  scale $M$
of the perturbative approach).
As `$n$' is the frequency of oscillation of $A_\mu$
with respect to $x^4$,
condition \eqref{eq:nonperturbative-condition} is similar to  condition
\eqref{eq:perturbative-condition} for the perturbative case.

Third, let us remark on the main improvements of our
present calculations compared with the earlier calculations
for $x^4$-independent background gauge fields.
Recall that the perturbative calculation here
used a generalized Pauli--Villars regularization
method with an extra infinite set of Pauli--Villars-type fields
$\psi_r$ (with regulator masses $M_r=M\,r^2$)
and maintains gauge invariance,
unlike the calculation of Ref.~\cite{Klinkhamer1999}
which used the standard Pauli--Villars regularization with a single set of regulator fields and a single regulator mass.
In the lattice calculation here, we have explicitly obtained the diagonalization operators $\mathcal{U}^{(n)}$ and have not used
an \textit{ad-hoc} phase fixing,
unlike the calculation of Ref.~\cite{KlinkhamerSchimmel2002}.

Fourth, let us try to understand heuristically why our new result
for $x^4$-dependent background gauge fields is similar to the
previous result for $x^4$-independent background gauge fields.
We see, from the result (\ref{cancellation of terms for odd N}),
that the anomalous terms arising from the positive frequency ($n>0$) are canceled by the terms arising from the negative frequency ($n<0$), so that only the term corresponding to $n=0$ contributes to the CPT violation,
which also has $m'=0$ according to \eqref{eq:det-n-equal-0}
[recall  \eqref{change in the effective action under cpt transformation}
for the definition of the Fourier modes $n$ and $m'$ entering
the change of the effective action under CPT].
This explains why, for the case of $x^4$-dependent background
gauge fields,
we have obtained a result similar to the one for the case of
$x^4$-independent gauge fields~\cite{Klinkhamer1999,KlinkhamerSchimmel2002}.
Indeed, compare \eqref{eq:change-under-CPT-odd-N} from the present paper,
with a unitary operator depending on
$x^4$-dependent gauge fields and a sum over $(x^1,\,x^2,\,x^3,\,x^4)$
in the determinant,
to (5.35) from Ref.~\cite{KlinkhamerSchimmel2002},
with essentially the same unitary operator depending on
$x^4$-independent gauge fields and only a sum over $(x^1,\,x^2,\,x^3)$
in the determinant.

Fifth, let us continue the heuristic discussion and
comment on the absence of $\partial_4\,A_i$ terms in our result.
We have calculated, in the perturbative approach,
the effective gauge-field action up to
two-point functions (second-order in the gauge field $A_\mu$).
In this approach, the CPT-anomalous terms are independent of the momentum in the fourth direction.
See, in particular, the discussion above \eqref{eq:T-of-pn-first},
where the $\rho_n$ term corresponds to
the position-space partial derivative $\partial_4$.
If we consider the non-Abelian gauge theory,
the CPT-anomalous terms will involve
three-point functions (third-order in the gauge field $A_\mu$).
There is then the possibility that the CPT-anomalous terms involving $\partial_4$ will not vanish by symmetry reasons.
For the continuum limit of the lattice calculation, we have also considered
only Abelian gauge fields and have expanded only up to
the two-point function $\Gamma_2[A]$ (second-order in the coupling constant $e$). For the non-Abelian case, we expect to have higher-order contributions (notably $\Gamma_3[A]$),
which may, in principle, give rise to terms involving
the partial derivative $\partial_4$ acting
on the background gauge field.

Sixth, recall that finite-temperature field theory can be described by a
quantum field theory defined over a Euclidean spacetime with a
compactified coordinate~\cite{MontvayMunster1997}.
This Euclidean-path-integral formulation of finite-temperature
field theory has the
same manifold as our theory ($\mathbb{R}^3 \times S^1$),
with $S^1$ coordinate $x^4 \in [0,L]$.
The range of the compactified coordinate is determined by
$L = \beta$,
where $\beta$ is the inverse of the temperature $T$ (in units with
$k_B=1$). The discrete momentum components
of the fermion fields (Matsubara frequencies)
are given by $p_4 = (n + 1/2)\,2\pi/\beta$, with integers
$n = 0,\, \pm 1,\, \pm 2\, \ldots$.

In several recent articles (see, e.g.,
Refs.~\cite{Cervi-etal-2001,Mariz-etal-2005}
and references therein), calculations have been reported
of a radiatively-induced Chern--Simons-like term in
four-dimensional finite-temperature field theory.
This temperature-dependent induced Chern--Simons-like term
violates the Lorentz and CPT symmetries.

But compared to our calculation there are significant differences.
Most importantly,
the fermions of the finite-temperature calculations
have anti-periodic boundary conditions
(coming from the trace in the partition function of the
finite-temperature system and having anti-commuting fields),
whereas we assume a periodic spin structure
over the compact dimension.
In our calculation, the anomalous Chern-Simons-like term results
from the zero-momentum part of the fermions,
which would be absent for anti-periodic boundary conditions.

In addition, the finite-temperature calculations
have an explicit Lorentz-violating term in the
fermion sector with a constant $b_\mu$
(the induced Chern-Simons-like term is proportional
to this constant $b_\mu$),
whereas the Lorentz violation in our calculation
comes from the regulator fields.
Moreover, the fermions of the finite-temperature calculations
can have a mass $m$,
whereas the original chiral fermions of our calculation are
strictly massless.

As a final comment, we emphasize the importance
of maintaining microcausality,
also for the finite-temperature effective theory in the $T \to 0$ limit
(cf. Refs.~\cite{Cervi-etal-2001,AdamKlinkhamer2001plb}).

%%\newpage%%tmp
\section{Conclusion}
\label{sec:Conclusion}

For the appropriate setup of the physical system (Sec.~\ref{sec:Setup-problem}), we have established perturbatively
(Sec.~\ref{sec:Perturbative-approach})
the existence of a CPT anomaly for a background gauge field $A_\mu$
which depends on the compactified $x^4$ coordinate
and has a vanishing component $A_4$.
We have also performed a nonperturbative calculation with a
lattice regularization (Sec.~\ref{sec:Nonperturbative-approach}) and
have discussed the continuum limit of the lattice result.
The nonperturbative result \eqref{eq:Gamma-CS-like-nonperturbative} agrees
with the earlier result \eqref{eq:Gamma-anom-perturbative-Gamma-CS-like}
obtained via the perturbative approach.
(In principle, these results could have differed
by an odd-integer prefactor, because, as noted in
Refs.~\cite{Klinkhamer1999,KlinkhamerSchimmel2002}
and Sec.~\ref{subsubsec:Fixing the phases} here,
there is an ambiguity in the anomalous term due to
the freedom in defining the regularized theory.)

The fact that the perturbative and nonperturbative
results for the CPT anomaly essentially agree is reminiscent of the
Adler--Bardeen result for the triangle
anomaly~\cite{AdlerBardeen1969}.
In this respect, note that the CPT anomaly of the
perturbative calculation originates in the $m=0$ sector
of the vacuum-polarization kernel \eqref{eq:pimunu}
with a linearly-diverging one-loop Feynman diagram.
Still,  it needs to be verified that there arise
no further terms in the nonperturbative lattice calculation.

Having a possible anomalous origin of the local
Chern--Simons-like term \eqref{eq:Gamma-CS-like}
in the effective gauge-field action provides additional incentive
to study the phenomenology of the so-called
Maxwell--Chern--Simons (MCS) theory~\cite{CarrollFieldJackiw1990}.
This MCS theory contains, in the photonic sector, the
standard Maxwell term and the local Chern--Simons-like term.
The MCS theory can also be augmented by the addition of
the standard gauge-invariant kinetic term of a Dirac
spinor field (the electron-positron field).

This MCS theory appears in two
varieties: one variety is parity-violating and time-reversal-invariant
(corresponding to a timelike $x^4$ coordinate in our calculation)
and the other variety is parity-conserving and time-reversal-noninvariant
(corresponding to a spacelike $x^4$ coordinate in our calculation).
Now, it is clear that our calculation for a  timelike $x^4$ coordinate
would start from a theory with closed timelike loops and such a theory
is, most likely, inconsistent~\cite{Hawking1992}.
It has, indeed, been shown that the parity-violating
(and time-reversal-invariant) variety of
MCS theory is noncausal and nonunitary~\cite{AdamKlinkhamer2001npb}.
The parity-conserving (and time-reversal-noninvariant)
variety of MCS theory appears to be
well-behaved~\cite{AdamKlinkhamer2001npb} and displays some interesting
nonstandard effects such as
photon triple-splitting~\cite{AdamKlinkhamer2003,KaufholdKlinkhamer2006}
and vacuum Cherenkov radiation~\cite{KaufholdKlinkhamer2006,KaufholdKlinkhamer2007}.

%%\newpage


\vspace*{5mm}
\begin{thebibliography}{00}

\bibitem{Klinkhamer1999}
F.R.~Klinkhamer,
``A CPT anomaly,''
Nucl.\ Phys.\ B {\bf 578}, 277 (2000),
arXiv:hep-th/9912169.
%%CITATION = doi:10.1016/S0550-3213(00)00117-6;%%

\bibitem{KlinkhamerNishimura2000}
F.R.~Klinkhamer and J.~Nishimura,
``CPT anomaly in two-dimensional chiral $U(1)$ gauge theories,''
 Phys.\ Rev.\ D {\bf 63}, 097701 (2001),
arXiv:hep-th/0006154.
%%CITATION = doi:10.1103/PhysRevD.63.097701;%%


\bibitem{KlinkhamerSchimmel2002}
F.R.~Klinkhamer and J.~Schimmel,
  ``CPT anomaly: A rigorous result in four-dimensions,''
  Nucl.\ Phys.\ B {\bf 639}, 241 (2002),
arXiv:hep-th/0205038.
  %%CITATION = doi:10.1016/S0550-3213(02)00543-6;%%


\bibitem{Klinkhamer2005}
F.R.~Klinkhamer,
``Nontrivial spacetime topology, CPT violation, and photons,''
%in: \emph{CP Violation and the Flavour Puzzle:
%Symposium in Honour of Gustavo C. Branco},
%D. Emmanuel-Costa \emph{et al} (eds.),
%Krakow, Poligrafia Inspektoratu, 2005, pp. 157--191,
arXiv:hep-ph/0511030.
%%CITATION = HEP-PH/0511030;%%

\bibitem{ChadhaNielsen1982}
S.~Chadha and H.B.~Nielsen,
``Lorentz invariance as a low-energy phenomenon,''
Nucl.\ Phys.\ B {\bf 217}, 125  (1983).
%%CITATION = doi:10.1016/0550-3213(83)90081-0;%%


\bibitem{CarrollFieldJackiw1990}
 S.M.~Carroll, G.B.~Field, and R.~Jackiw,
``Limits on a Lorentz and parity violating modification of electrodynamics,''
Phys.\ Rev.\ D {\bf 41}, 1231 (1990).
  %%CITATION = doi:10.1103/PhysRevD.41.1231;%%

\bibitem{ColladayKostelecky1998}
D.~Colladay and V.A.~Kostelecky,
``Lorentz violating extension of the standard model,''
Phys.\ Rev.\ D {\bf 58}, 116002 (1998),
arXiv:hep-ph/9809521.
%%CITATION = doi:10.1103/PhysRevD.58.116002;%%

\bibitem{FrolovSlavnov1993}
S.A.~Frolov and A.A.~Slavnov,
  ``An invariant regularization of the Standard Model,''
  Phys.\ Lett.\ B {\bf 309}, 344 (1993).
%%CITATION = doi:10.1016/0370-2693(93)90943-C;%%


\bibitem{GinspargWilson1982}
P.H.~Ginsparg and K.G.~Wilson,
``A remnant of chiral symmetry on the lattice,''
Phys.\ Rev.\ D {\bf 25}, 2649 (1982).
%%CITATION = doi:10.1103/PhysRevD.25.2649;%%


\bibitem{Neuberger1998a}
H.~Neuberger,
  ``Exactly massless quarks on the lattice,''
  Phys.\ Lett.\ B {\bf 417}, 141 (1998),
arXiv:hep-lat/9707022.
%%CITATION = doi:10.1016/S0370-2693(97)01368-3;%%

\bibitem{Neuberger1998b}
  H.~Neuberger,
  ``More about exactly massless quarks on the lattice,''
  Phys.\ Lett.\ B {\bf 427}, 353 (1998),
arXiv:hep-lat/9801031.
  %%CITATION = doi:10.1016/S0370-2693(98)00355-4;%%


\bibitem{Luscher1998a}
 M.~L\"{u}scher,
  ``Exact chiral symmetry on the lattice and the Ginsparg--Wilson relation,''
  Phys.\ Lett.\ B {\bf 428}, 342 (1998),
arXiv:hep-lat/9802011.
%%CITATION = doi:10.1016/S0370-2693(98)00423-7;%%


\bibitem{Luscher1998b}
  M.~L\"{u}scher,
  ``Abelian chiral gauge theories on the lattice with exact gauge invariance,''
  Nucl.\ Phys.\ B {\bf 549}, 295 (1999),
arXiv:hep-lat/9811032.
  %%CITATION = doi:10.1016/S0550-3213(99)00115-7;%%


\bibitem{AitchisonFoscoZuk1993}
I.J.R.~Aitchison, C.D.~Fosco, and J.A.~Zuk,
``On the temperature dependence of the induced Chern--Simons term in (2+1)-dimensions,''
Phys.\ Rev.\ D {\bf 48}, 5895 (1993).
%%CITATION = doi:10.1103/PhysRevD.48.5895;%%

\bibitem{BirrellDavies1982}
N.D.~Birrell and P.C.W.~Davies,
\emph{Quantum Fields in Curved Space},
Cambridge University Press, 1982.
 %%CITATION = doi:10.1017/CBO9780511622632;%%


\bibitem{ChernSimons1974}
S.S.~Chern and J.~Simons,
``Characteristic forms and geometric invariants,''
Annals Math.\  {\bf 99}, 48 (1974).
%%CITATION = doi:10.2307/1971013;%%

\bibitem{MontvayMunster1997}%%{MontvayMunster1997}
I.~Montvay and G.~M\"{u}nster,
 \emph{Quantum Fields on a Lattice},
Cambridge University Press, 1997.
%%CITATION = INSPIRE-378182;%%

\bibitem{CosteLuscher1989}
 A.~Coste and M.~L\"{u}scher,
  ``Parity anomaly and fermion boson transmutation in three-dimensional lattice QED,''
  Nucl.\ Phys.\ B {\bf 323}, 631 (1989).
%%CITATION = doi:10.1016/0550-3213(89)90127-2;%%

\bibitem{Cervi-etal-2001}
L.~Cervi, L.~Griguolo, and D.~Seminara,
``The structure of radiatively induced Lorentz and CPT violation
in QED at finite temperature,''
Phys.\ Rev.\ D {\bf 64}, 105003 (2001),
arXiv:hep-th/0104022.
%%CITATION = doi:10.1103/PhysRevD.64.105003;%%


\bibitem{Mariz-etal-2005}
T.~Mariz, J.R.~Nascimento, E.~Passos, R.F.~Ribeiro, and F.A.~Brito,
``A remark on Lorentz violation at finite temperature,''
JHEP {\bf 0510}, 019 (2005),
arXiv:hep-th/0509008.
%%CITATION = doi:10.1088/1126-6708/2005/10/019;%%

\bibitem{AdamKlinkhamer2001plb}
C.~Adam and F.R.~Klinkhamer,
``Causality and radiatively induced CPT violation,''
Phys.\ Lett.\ B {\bf 513}, 245 (2001),
arXiv:hep-th/0105037.
%%CITATION = doi:10.1016/S0370-2693(01)00678-5;%%

\bibitem{AdlerBardeen1969}
S.L.~Adler and W.A.~Bardeen,
``Absence of higher order corrections in the anomalous axial vector divergence equation,''
Phys. Rev.  {\bf 182}, 1517 (1969).
%%CITATION = doi:10.1103/PhysRev.182.1517;%%

\bibitem{Hawking1992}
S.W.~Hawking,
``The chronology protection conjecture,''
Phys.\ Rev.\ D {\bf 46}, 603  (1992).
%%CITATION = doi:10.1103/PhysRevD.46.603;%%

\bibitem{AdamKlinkhamer2001npb}
C.~Adam and F.R.~Klinkhamer,
  ``Causality and CPT violation from an Abelian Chern--Simons-like term,''
Nucl.\ Phys.\ B {\bf 607}, 247 (2001),
arXiv:hep-ph/0101087.
  %%CITATION = doi:10.1016/S0550-3213(01)00161-4;%%

\bibitem{AdamKlinkhamer2003}
C.~Adam and F.R.~Klinkhamer,
``Photon decay in a CPT violating extension of quantum electrodynamics,''
Nucl. Phys. B {\bf 657}, 214 (2003),
arXiv:hep-th/0212028.
%%CITATION = doi:10.1016/S0550-3213(03)00143-3;%%

\bibitem{KaufholdKlinkhamer2006}
C.~Kaufhold and F.R.~Klinkhamer,
``Vacuum Cherenkov radiation and photon triple-splitting in a Lorentz-noninvariant extension of quantum electrodynamics,''
Nucl. Phys. B {\bf 734}, 1 (2006),
arXiv:hep-th/0508074.
%%CITATION = doi:10.1016/j.nuclphysb.2005.11.001;%%


\bibitem{KaufholdKlinkhamer2007}
C.~Kaufhold and F.R.~Klinkhamer,
``Vacuum Cherenkov radiation in spacelike Maxwell--Chern--Simons theory,''
Phys.\ Rev.\ D {\bf 76}, 025024 (2007),
arXiv:0704.3255.  %% [hep-th]].
%%CITATION = doi:10.1103/PhysRevD.76.025024;%%


\end{thebibliography}
\end{document}